\documentclass{aa}
\usepackage{float}
\usepackage{graphicx,natbib,txfonts,color}
\usepackage{minipage-marginpar}
\usepackage{amssymb, amsmath} 
\usepackage[separate-uncertainty=true]{siunitx}
\usepackage{multicol}
\usepackage{placeins}
\usepackage{hyperref}
\usepackage{pdflscape}
\usepackage{rotating}

\bibpunct{(}{)}{;}{a}{}{,}

\newcommand{\mjup}{\mathrm{M}_{\text{Jup}}}
\newcommand{\rjup}{\mathrm{R}_{\text{Jup}}}

\newcommand{\Zsun}{\mathrm{Z}_{\odot}}
\newcommand{\rsun}{\mathrm{R}_{\odot}}

\newcommand{\rplanet}{\mathrm{R}_{\text{pl}}}

\newcommand{\AU}{\text{AU}}

\newcommand{\Kzz}{\mathrm{K}_{\rm zz}}
\newcommand{\Teq}{\mathrm{T}_\text{eq}}
\newcommand{\Teff}{\mathrm{T}_\text{eff}}

\newcommand{\cmsquareds}{\, \text{cm}^2/\text{s}}
\newcommand{\cmspersquared}{\, \text{cm}/\text{s}^2}

\newcommand{\um}{\, \mu \text{m}}
\newcommand{\mbar}{\, \text{mbar}}
\newcommand{\mubar}{\, \mu\text{bar}}
\newcommand{\Tint}{\, \mathrm{T}_{\text{int}}}

\newcommand{\Pbase}{\, \text{P}_{\text{cloud}}}
\newcommand{\Xbase}{\, \text{X}_{\text{cloud}}}
\newcommand{\fsed}{\, \text{f}_\text{sed}}
\newcommand{\sigmag}{\, \sigma_\text{g}}
\newcommand{\Kzzcloud}{\text{K}^\textbf{\rm (cloud)}_\text{zz}}
\newcommand{\kappaWelbanks}{\, \kappa_{\rm opac}}
\newcommand{\kappascatt}{\, \kappa_{\rm scatt}}
\newcommand{\gammascatt}{\, \gamma_{\rm scatt}}
\newcommand{\xsigma}{\, x_\sigma}

\newcommand{\offsetHST}{\Delta_\text{HST}}
\newcommand{\offsetNIR}{\Delta_\text{NIR}}

\newcommand{\Pref}{\text{P}_\text{ref}}

\usepackage[version=4]{mhchem}

\begin{document}
\let\linenumbers\nolinenumbers
\nolinenumbers

\title{Reliability of 1D radiative-convective photochemical-equilibrium retrievals on transit spectra of WASP-107b}

\author{T. Konings  \inst{1},
	L. Heinke \inst{1, 2, 3},
	R. Baeyens \inst{4},
	K. Hakim \inst{1, 5},
	V. Christiaens \inst{1, 6},
\and L. Decin \inst{1}
}


\institute{${}^1$Institute of Astronomy, KU Leuven, Celestijnenlaan 200D, 3001 Leuven, Belgium \\
	${}^2$Centre for Exoplanet Science, University of Edinburgh, Edinburgh, EH9 3FD, UK \\
	${}^3$School of GeoSciences, University of Edinburgh, Edinburgh, EH9 3FF, UK \\
	${}^4$Anton Pannekoek Institute for Astronomy, University of Amsterdam, Science Park 904, 1098 XH Amsterdam, The Netherlands \\
	${}^5$Royal Observatory of Belgium, Ringlaan 3, 1180 Brussels, Belgium \\
	${}^6$University of Liège, STAR Institute, Allée du Six Août 19C, 4000 Liège, Belgium
	}

\date{Accepted 13 July 2025}

\authorrunning{T. Konings et al.}
\titlerunning{1D-RCPE retrievals of WASP-107b}

\abstract
{
WASP-107b has been observed in unprecedented detail with the James Webb Space Telescope. These observations suggest that it has a metal-rich and carbon-deprived atmosphere with an extremely hot interior based on detections of \ce{SO2}, \ce{H2O}, \ce{CO2}, \ce{CO}, \ce{NH3}, and \ce{CH4}.
}
{
In this paper, we aim to determine the reliability of a 1D radiative-convective photochemical-equilibrium (1D-RCPE) retrieval method in inferring atmospheric properties of WASP-107b.
We aim to explore its sensitivity to modelling assumptions and different cloud parametrizations, and investigate the data information content.
Additionally, we aim to characterize chemical trends and map dominant pathways to develop a comprehensive understanding of the 1D-RCPE model grid before running the retrievals.
}
{
We built a grid of radiative-convective balanced pressure-temperature profiles and 1D photochemical equilibrated models, which cover a range of metallicities (Z), carbon-to-oxygen ratios (C/O), intrinsic temperatures ($\Tint$), and eddy diffusion coefficients ($\Kzz$).
We adopted a nested sampling algorithm within a Bayesian framework to estimate model parameters from previously analysed transit observations of WASP-107b discontinuously covering $1.1$ to $12.2 \um$.
}
{
Our model grid reproduces established chemical trends such as the dependence of \ce{SO2} production on metallicity and demonstrates that mixing-induced quenching at high $\Tint$ reduces the bulk \ce{CH4} and \ce{NH3} content.
We obtain good fits with our 1D-RCPE retrievals that are mostly based on a few molecular features of \ce{H2O}, \ce{CO2}, \ce{SO2}, and \ce{CH4}, but find no substantial contribution of \ce{NH3}.
We find that the degeneracy between metallicity, cloud pressure, and a model offset is broken by the presence of strong \ce{SO2} features,
confirming that \ce{SO2} is a robust metallicity indicator.
We systematically retrieve sub-solar C/O based on the relative amplitude of a strong \ce{CO2} feature
with respect to the broad band of \ce{H2O},
which is sensitive to a wavelength-dependent scattering slope.
We find that high-altitude clouds obscure the \ce{CH4}-rich layers, preventing the retrievals from constraining $\Tint$, but that
higher values of $\Kzz$ 
can transport material above the cloud deck, allowing a fit of the \ce{CH4} feature.
However, $\Tint$ and $\Kzz$ can vary substantially between retrievals depending on the adopted cloud parametrization.
}
{
We conclude that the 1D-RCPE retrieval method can provide useful insights if the underlying grid of forward models is well understood.
We find that WASP-107b's atmosphere is enriched in metals ($3-5 \, \Zsun$) and carbon-deprived (C/O $\lesssim$ 0.20).
However, we lack robust constraints on the intrinsic temperature and vertical mixing strength.
}


\maketitle


\section{Introduction}\label{sec: intro}
WASP-107b is a transiting exoplanet orbiting a K7V-type host star
at a distance of $\num{0.055} \, \AU$,
resulting in an equilibrium temperature of ${\sim}740$ K \citep{Anderson2017discoveryofW107b}.
With a mass of 0.12 $\mjup$ and a radius of 0.90 $\rjup$ \citep{Piaulet2021densityW107b}, its extremely low gravity and large scale height
make it well suited for atmospheric characterization.
A series of transit observations have revealed the structure of WASP-107b's deeper atmosphere in unprecedented detail.
Observations with the Hubble Space Telescope (HST) in the near-infrared show water vapour features, compressed by high-altitude clouds \citep{Kreidberg2018waterw107b}. 
Further observations with James Webb Space Telescope's (JWST) MIRI instrument (LRS) were obtained by \citet{Dyrek2023so2SilicateClouds}, who further constrained the type of aerosols to silicate cloud particles (i.e. \ce{MgSiO3}, \ce{SiO2}, and \ce{SiO}).
Additionally, the detection of \ce{SO2} is indicative of photochemistry in a metal-enriched atmosphere \citep{Tsai2023PhotochemicallyproducedSO2, Dyrek2023so2SilicateClouds}.
\newline

\noindent Neither dataset shows proof of \ce{CH4}, despite predictions from equilibrium chemistry showing that methane should be abundant at temperatures below ${\sim}1000$ K \citep{Moses2011-Disequilibrium}.
Vertical mixing drives the atmosphere out of equilibrium and can quench \ce{CH4} in the deep atmosphere, thereby fixing the methane content in higher layers to the volume mixing ratio (VMR) at the quench point.
In addition to possible carbon depletion by a sub-solar carbon-to-oxygen ratio, \citet{Dyrek2023so2SilicateClouds} show that the combination of a high intrinsic temperature and strong vertical mixing reduces the \ce{CH4} content to undetectable levels without affecting the formation of \ce{SO2} in the upper layers (see their Extended Data - Figure 3).
\newline

\noindent More recently, \citet{Welbanks2024-High-InternalHeatFlux} analysed JWST NIRCam spectra in combination with previous datasets and managed to detect and constrain the abundances of \ce{H2O}, \ce{SO2}, \ce{CO}, \ce{CO2}, \ce{NH3}, and even traces of \ce{CH4}.
These results were accompanied by the publication of \citet{Sing2024-WarmNeptunes}, who used a JWST NIRSpec spectrum to detect \ce{H2O}, \ce{SO2}, \ce{CO}, \ce{CO2}, and \ce{CH4}.
Both of the above studies conclude that the depleted \ce{CH4} is the result of a high intrinsic temperature (> 350 K) and atmospheric mixing causing deep quenching.
Additionally, \citet{Welbanks2024-High-InternalHeatFlux} attribute this hot interior to eccentricity-driven tidal heating, which had previously been identified as a potential explanation for the inflated radius of WASP-107b
\citep{Thorngren2019TheIntrinsicTemperature, Fortney2020BeyondEq, Yu2024AreWasp107blikeSystems}.
\\

\noindent
Atmospheric retrievals aim to infer basic properties
from observations by navigating a large, degenerate parameter space using parametrized, easy-to-compute models and Bayesian inference techniques \citep{Madhusudhan2018AtmosphericRetrieval}.
However, such simplifications can lead to biases for inferred parameters, which have been documented and investigated in the past \citep[e.g.][]{BennekeAndSaeger2012AtmosphericRetrieval, Griffith2014Disentagnling,  Line2016InfluenceOfNonuniform, Heng2017theory, Fisher2018Retrieval, Welbanks2019degeneracies,MacDonald2020-WhyIsitSoColdinHere, Novais2025MNRAS.tmp..381N}.
Alternatively, one can construct more complex forward models that take into account, for example, chemical kinetics \citep[e.g. in][]{Dyrek2023so2SilicateClouds}, radiative-convective equilibrium, cloud microphysics, and fluid dynamics.
A major downside of forward models is the large computational cost, which makes them unsuited for retrievals.
Combining the best of both worlds, \cite{Bell2023MethaneThroughouttheatmosphere} introduced a 1D radiative-convective photochemical-equilibrium (1D-RCPE) retrieval \citep{Beatty2024SulfurDioxideGJ3470, Fu2024H2S_HD189733b, Welbanks2024-High-InternalHeatFlux}.
By pre-computing a grid of physics-informed forward models, and allowing for interpolation in between, 1D-RCPE retrievals can leverage the strength of Bayesian inference with on-the-fly computation of low-cost forward models.
This technique was applied to WASP-107b by \citet{Welbanks2024-High-InternalHeatFlux}, who deduced a high intrinsic temperature based on the \ce{CH4} deficiency.
\\

\noindent Given WASP-107b's inflated radius and low density, the atmospheric elemental composition and intrinsic temperature can help to unravel properties such as the planet's core mass, and formation history \citep{Madhusudhan2016-ExoplanetaryAtmospheres:ChemistryFormationandHabitability, Anderson2017discoveryofW107b, Piaulet2021densityW107b,Welbanks2024-High-InternalHeatFlux, Sing2024-WarmNeptunes}.
To achieve this, it is crucial to deduce the atmospheric properties from transit spectra as accurately as possible.
Various methods have been adopted in the above studies, which lead to different results.
More specifically for WASP-107b, reported values for atmospheric metallicity range from $\gtrsim 6 \, \Zsun$ \citep{Dyrek2023so2SilicateClouds} up to $43 \pm 8\, \Zsun$ \citep{Sing2024-WarmNeptunes}, and for carbon-to-oxygen ratios from $0.51_{-0.21}^{+0.27}$ (solar) \citep{Sing2024-WarmNeptunes} down to $0.33_{-0.05}^{+0.06}$ \citep{Welbanks2024-High-InternalHeatFlux}.
Additionally, multiple studies on WASP-107b point towards a high intrinsic temperature and deep quenching based on the lack of strong \ce{CH4} features.
\\

\noindent In this paper, we aim to assess how well 1D-RCPE retrievals can reliably infer atmospheric properties such as metallicity (Z), carbon-to-oxygen ratio (C/O), intrinsic temperature ($\Tint$), and eddy diffusion coefficient ($\Kzz$) for WASP-107b.
More specifically, we explore the sensitivity of our results to different cloud parametrizations, and investigate the information content of the available observations.
A secondary goal of this paper is to consistently analyse the relevant chemical pathways and summarize the chemical trends with respect to the four grid parameters (Z, C/O, $\Tint$, and $\Kzz$).
\\

%
%
\noindent In Sect. \ref{sec: methods}, we describe the methods and numerical codes used in this study.
We start by constructing a grid of 900 1D-RCPE models and analyse the results
in Sect. \ref{sec: results chemistry}. 
We then proceed by running 1D-RCPE retrievals with various set-ups in Sect. \ref{sec: results grid fit}.
Finally, in Sect. \ref{sec: discussion}, we discuss how our results affect previously inferred properties of WASP-107b before concluding in Sect. \ref{sec: conclusion}.

\section{Methods} \label{sec: methods}

We used a number of different codes to build and analyse our grid of 1D-RCPE models.
First, we outline the design of our model grid (Sect. \ref{subsec: grid design}).
This is followed by description of our model set-up for the pressure-temperature profiles with \textsc{PICASO} (Sect. \ref{subsec: picaso}) and the chemical kinetics simulations with VULCAN (Sect. \ref{subsec: vulcan}).
We analysed the resulting grid of 900 1D-RCPE models with Dijkstra's algorithm (Sect. \ref{subsec: Dijkstra}).
To generate synthetic transit spectra, we used petitRADTRANS (Sect. \ref{subsec: petitRADTRANS}) with various cloud parametrizations (Sect. \ref{subsec: cloudparams}).
Finally, we fitted the 1D-RCPE model grid to different datasets (Sect. \ref{subsec: Data Sources and model offsets}) with UltraNest (Sect. \ref{subsec: Ultranest}).

\subsection{Grid design} \label{subsec: grid design}

We varied four parameters in our grid of forward models: Z, C/O, $\Tint$, and $\Kzz$. 
We ran atmospheric models with metallicities 1, 2, 3, 5, 7, 10, 15, 20, and 30 $Z_\odot$ \citep{Asplund2009-TheChemical}. 
We discarded sub-solar values based on previous findings and the detection of \ce{SO2}, which points to metal enrichment \citep{Kreidberg2018waterw107b, Dyrek2023so2SilicateClouds}. 
The upper limit was chosen based on the interior structure models that estimate the planet's maximal bulk metallicity \citep{Kreidberg2018waterw107b,Thorngren2019ConnectingGiantPlanetAtmosphereandInteriorModeling}.
We explored C/O ratios of 0.05, 0.11, 0.23, and 0.46, corresponding to roughly 10\%, 25\%, 50\%, and 100\% of the solar value, respectively \citep{Lodders2010ASSP...16..379L}, in agreement the lower boundary suggested by planet formation models \citep{Olberg2011EffectofSnowlines, Mordasini2016Imprintofexoplanetsformationhistory, Espinoza2017MetalEnrichmentLeadstolowatmospheric}.
We considered intrinsic temperatures of 100, 200, 300, 400, and 500 K.
The lower values correspond to the predictions where only deposition of stellar radiation is considered \citep{Thorngren2019TheIntrinsicTemperature}, while higher values correspond to increasing effects of tidal heating \citep{Leconte2010IsTidalHeatingSufficient, Fortney2020BeyondEq, Millholland2020TidalInflationReconciles}.
We explored eddy diffusion coefficients of $10^6$, $10^7$, $10^8$, $10^9$, and $10^{10} \cmsquareds$, corresponding to atmospheres with and without quenching.
Combined, this results in a grid of 900 forward models.
Throughout the grid, we used parameters that are representative for the WASP-107(b) system.
For WASP-107b, these include a surface gravity of $~270\cmspersquared$, radius of $0.94\, \rjup$, 
mass of $0.10\, \mjup$, orbital separation of $0.055\, \AU$, and equilibrium temperature of 740 K.
For its host star, we adopted $\Teff^\star = 4430 \, \text{K}$, a radius of $0.66 \, \rsun$, and $\log g$ (cgs) $\simeq 4.6$ \citep{Piaulet2021densityW107b}.

\subsection{Radiative-convective equilibrium: \textsc{PICASO}} \label{subsec: picaso}

We used the open-source 1D radiative-convective code \textsc{PICASO} \citep{Mukherjee2023PICASO} to compute 180 temperature-pressure profiles of the atmosphere of WASP-107b, each with a different metallicity, carbon-to-oxygen ratio, and intrinsic temperature. 
We ran \textsc{PICASO} with 91 pressure log-spaced levels between $300$ and $10^{-7}$ bar.
The code uses pre-mixed correlated-k opacities that are calculated under the assumption of equilibrium chemistry \citep{lupu_2023_7542068, Mukherjee2023PICASO}. 
As is listed in \citet{lupu_2023_7542068}, these include
\ce{C2H2} \citep{Wilzewski2016JQSRT.168..193W, HITRAN2012Rothman2013HITRAN},
\ce{C2H4} \citep{HITRAN2012Rothman2013HITRAN, M21-Marley2021ApJ...920...85M},
\ce{C2H6} \citep{HITRAN2012Rothman2013HITRAN, M21-Marley2021ApJ...920...85M},
\ce{CH4} \citep{Yurchenko2013JMoSp.291...69Y, Yurchenko2014MNRAS.440.1649Y, Pine1992JChPh..97..773P, STDS-Wenger1998JQSRT..59..471W},
\ce{CO} \citep{HITEMP2010-Rothman2010JQSRT.111.2139R, HITRAN2016Gordon2017JQSRT.203....3G, Li2015ApJS..216...15L},
\ce{CO2} \citep{HUANG2014134},
\ce{CrH} \citep{Burrows2002ApJ...577..986B},
\ce{Fe} \citep{VALD3-Ryabchikova2015PhyS...90e4005R, OBrian1991JOSAB...8.1185O, Fuhr1988JPCRD..17S....F, Bard1991A&A...248..315B, Bard1994A&A...282.1014B},
\ce{FeH} \citep{Dulick2003ApJ...594..651D, Hargreaves2010AJ....140..919H},
\ce{H2} \citep{HITRAN2016Gordon2017JQSRT.203....3G},
\ce{H3+} \citep{Mizus2017MNRAS.468.1717M},
\ce{H2O} \cite{Polyansky2018H2PExomolLinelist, Barton2017JQSRT.203..490B},
\ce{H2S} \citep{Azzam2016ExoMolH2S, HITRAN2012Rothman2013HITRAN, M21-Marley2021ApJ...920...85M},
\ce{HCN} \citep{Harris2006MNRAS.367..400H, Barber2014MNRAS.437.1828B, HITRAN2020Gordon2022JQSRT.27707949G},
\ce{LiCl} \citep{Bittner2018ApJS..236...46B},
\ce{LiF} \citep{Bittner2018ApJS..236...46B},
\ce{LiH} \citep{Coppola2011MNRAS.415..487C},
\ce{MgH} \citep{GharibNezhad2013MNRAS.432.2043G, Yadin2012MNRAS.425...34Y, GN21a-Gharib-Nezhad2021ApJS..254...34G},
\ce{N2} \citep{HITRAN2012Rothman2013HITRAN, M21-Marley2021ApJ...920...85M},
\ce{NH3} \citep{Yurchenko2011MNRAS.413.1828Y, Wilzewski2016JQSRT.168..193W, M21-Marley2021ApJ...920...85M},
\ce{OCS} \citep{HITRAN2016Gordon2017JQSRT.203....3G},
\ce{PH3} \citep{Sousa-Silva2014JQSRT.142...66S, M21-Marley2021ApJ...920...85M},
\ce{SiO} \citep{Barton2013MNRAS.434.1469B, Kurucz1992RMxAA..23...45K, M21-Marley2021ApJ...920...85M},
\ce{TiO} \citep{McKemmish2019MNRAS.488.2836M, GN21a-Gharib-Nezhad2021ApJS..254...34G},
and \ce{VO} \citep{McKemmish2019MNRAS.488.2836M, GN21a-Gharib-Nezhad2021ApJS..254...34G, M21-Marley2021ApJ...920...85M},
in addition to alkali metals (Li, Na, K, Rb, Cs) \citep{VALD3-Ryabchikova2015PhyS...90e4005R, Allard2019A&A...628A.120A, Allard2016A&A...589A..21A, Allard2007A&A...474L..21A, Allard2007EPJD...44..507A}.
For a planet such as WASP-107b, \ce{H2O} and potentially \ce{CH4} are expected to be the dominant infrared absorbers \citep{Molliere2015-ModelAtmospheresofIrradiated}. 
\newline

\begin{figure}
	\resizebox{\hsize}{!}{\includegraphics{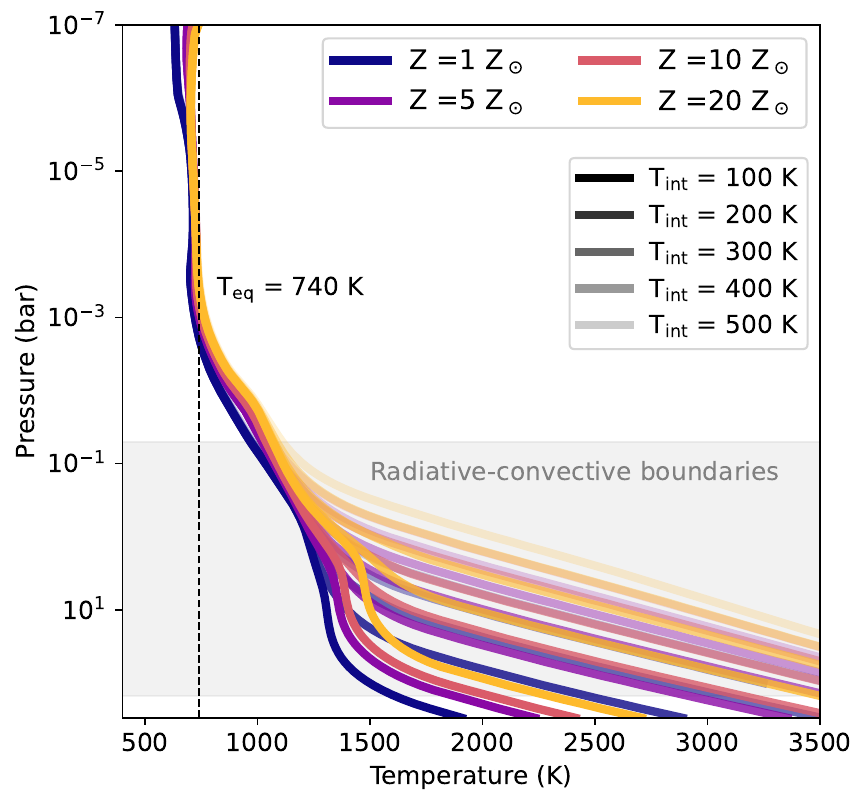}}
	\caption{Selection of the 180 computed PT profiles with \textsc{PICASO} for $\text{C/O} = 0.46$. The line transparency corresponds to intrinsic temperature, from a full line for $\Tint = 100$ K to almost transparent for $\Tint = 500$ K. The shaded grey region indicates layers where the radiative-convective boundaries are located in all 180 PT profiles.
	}
	\label{fig: temperatures_grid}
\end{figure}

\noindent
In order to converge to a solution with a smooth transition between the radiative and convective layers, \textsc{PICASO} requires an initial guess of the pressure level of the radiative-convective boundary zone (RCB).
As the RCB is highly sensitive to $\Tint$ \citep{Sarkis2021Evidenceofthreemechanisms}, we varied the RCB pressure level for each model with different intrinsic temperatures to ensure said convergence.
The pre-computed opacity mixtures in \textsc{PICASO} do not include metallicities $7$ and $15 \, Z_{\odot}$.
Therefore, we linearly interpolated these PT structures from neighbouring metallicities ($5$ to $10 \, Z_{\odot}$, and $10$ to $20 \, Z_{\odot}$, respectively).
For models with C/O = 0.05, we used the same computation as C/O = 0.11 as such low carbon-to-oxygen ratios are not available in the opacity data.
Finally, we chose to compute day-side averages by setting r$_{\rm st} = 1 $, which controls the contribution of stellar radiation to the net radiative flux in every atmospheric layer \citep[see Eq. 20 in][]{Mukherjee2023PICASO},
so that the temperature in the upper, quasi-isothermal layers matches the reported values in \citet{Dyrek2023so2SilicateClouds}.
We further discuss this assumption in Sect. \ref{sec: discussion}.
Fig. \ref{fig: temperatures_grid} shows a selection of the computed PT profiles with \textsc{PICASO}.

\subsection{(Photo)-chemical modelling: \textsc{VULCAN}} \label{subsec: vulcan}

We used the open-source chemical kinetics code VULCAN \citep{Tsaj2017-VULCANaopensource, Tsai2021ComparativeStudy} to model the molecular composition of WASP-107b.
We used the default chemical network \textsc{sncho\_photo\_network} that includes 89 species composed of H, C, O, N, S, 
coupled by 1030 forward and backward thermochemical reactions.
Vertical mixing was included via diffusion, with accompanying eddy diffusion coefficient $\Kzz$. 
We used a constant $\Kzz$ profile for each model, with the values specified in Sect. \ref{subsec: grid design}, to reduce complexity and maintain good interpretability of our results.
\\

\noindent
Photochemistry was included by way of 60 photodissociation reactions for the main UV absorbing species in \ce{H2}-dominated atmospheres.
We used a spectral energy distribution (SED) of HD 85512, as a proxy for host star WASP-107.
HD 85512 is a K7V star with a similar effective temperature as WASP-107, and its SED is available in the MUSCLES database \citep{France2016-TheMUSCLESTreasurySurveyIMotivationandOverview, Youngblood2016-TheMUSCLESTreasurySurveyII, Loyd2016-MUSCLESTreasurySruveyIII}.
Note that the SEDs of stars with similar spectral types are not necessarily comparable, as high-energy (X-ray, UV) radiation fluxes are linked to stellar age, rotation period and overall magnetic activity, rather than photospheric emission.
With a rotation period of $17 \pm 1$ days \citep{Anderson2017discoveryofW107b} and age of $3.4 \pm 0.7$ Gyr \citep{Piaulet2021densityW107b}, WASP-107 is expected to be more active compared to HD 85512, which has a ${\sim}47$ day rotation period and a resulting age estimate of $5.6 \pm 0.1$ Gyr \citep{Tsantaki2013-Derivingprecise}.
However, observations in the NUV and X-ray have shown that WASP-107 is slightly less active than HD 85512, although this difference is less than a factor of two in the UV so that HD 85512 remains a good proxy for WASP-107 \citep{Dyrek2023so2SilicateClouds}.
\\

\noindent
To initialise \textsc{VULCAN}, we used 150 log-spaced vertical layers between $10$ and $10^{-7}$ bar for models with high $\Tint$  to ensure convergence in the deep layers, while this upper pressure boundary can be extended to $300$ bar for low $\Tint$ to capture the quenching points.
To fix individual convergence issues, we lowered parameters $\textsc{atol}$ and/or $\textsc{rtol}$ (absolute and relative tolerance thresholds for errors during the numerical integration), or systematically decreased the pressure boundary at the interior while making sure the quench points of key species (i.e. \ce{CH4} and \ce{NH3}) are captured in the model.
Otherwise, we used the default set-up of VULCAN.
The above adjustments were especially necessary for models with low $\Kzz$, where the quench points are located much higher than their high-$\Kzz$ counterparts.
\\

\noindent The elemental mixture was adjusted based on the given combination of Z and C/O, before the simulation is started from chemical equilibrium abundances.
We first scaled the solar elemental mixture \citep{Asplund2009-TheChemical} with the Z, before lowering the carbon abundance to the requested C/O.
We did not include processes such as condensation, ion chemistry, or atmospheric escape.
Finally, we highlight that VULCAN assumes a fixed thermal background during the numerical integration. Studies have shown that the introduction of disequilibrium chemistry can lead to temperature differences up to 100 K in the PT profile \citep{Drummond2016-effectsofconsistent, Agundez2025arXiv250611658A}. 

\subsection{Chemical pathway analysis: Dijkstra's algorithm} \label{subsec: Dijkstra}

Alongside of chemical abundances, we are interested in the reaction rates in the steady-state solution of \textsc{VULCAN}.
This enables us to analyse the chemical pathways that govern the production and destruction of key species in our model.
In particular we are interested in the dominant or fastest chemical pathways, which translates to finding the shortest chain of inverse reaction rates that connect two molecules.
This is equivalent to finding the shortest path in graph theory, where the atmospheric constituents are nodes and reactions the connecting edges.
Therefore, we implement Dijkstra's algorithm \citep{dijkastra1959note}, following the implementation outlined in Appendix B of \citet{Tsai2018-TowardConsistentModeling}.

\subsection{Radiative transfer: petitRADTRANS} \label{subsec: petitRADTRANS}

We used the open-source radiative transfer code petitRADTRANS (version 2.7.7) \citep{Molliere2019-petitRADTRANS} to compute synthetic transmission spectra from the chemical and radiative-convective models.
We included correlated-k line absorption opacities of 
\ce{SO2} \citep{Underwood2016SO2linelist},
\ce{H2S} \citep{Azzam2016ExoMolH2S},
\ce{CO2} \citep{Yurchenko2020ExoMolCO2},
\ce{CO} \citep{Kurucz1993CO, HITEMP2010-Rothman2010JQSRT.111.2139R},
\ce{NH3} \citep{Coles2019ExoMolNH3},
\ce{CH4} \citep{Yurchenko2017ExoMolCH4}, 
and \ce{H2O} \citep{Polyansky2018H2PExomolLinelist} available via ExoMolOP \citep{Chubb2021ExoMolOPdatabase} and HITEMP \citep{HITEMP2010-Rothman2010JQSRT.111.2139R}.
To reduce computational cost, we left out line absorption opacity of \ce{SO}, \ce{C2H2}, \ce{OH}, and \ce{HCN} as we found no substantial contribution to the spectra of our grid.
Furthermore, we included Rayleigh scattering opacity of \ce{H2} \citep{Dalgarno1962-H2scattering} and \ce{He} \citep{Chan1965-HeScattering}, and continuum opacity from collision induced absorption (CIA) by \ce{H2}-\ce{H2} and \ce{H2}-\ce{He} \citep{Borysow1988-CIA, Borysow1989-CIA, Borysow1989-CIA_2, Borysow2001-CIA, Borysow2002-CIA, Richard2012-CIA}.

\subsection{Cloud opacity parametrizations} \label{subsec: cloudparams}
We considered different types of clouds when post-processing our grid of models to synthetic spectra. 
First, we considered a grey opacity source based at a certain pressure, $\Pbase$, which effectively blocks the layers below from forming spectral features.
Second, we considered non-grey clouds by adding an additional opacity source to the radiative transfer models.
Following \citet{Dyrek2023so2SilicateClouds}, we added silicates consisting of crystalline
\ce{MgSiO3}, 
\ce{SiO2}, 
or amorphous \ce{SiO}
cloud particles that are irregularly shaped, for which the opacity was computed with the DHS (distribution of hollow spheres) method \citep{Min2005Modelingopticalproperties}.
Due to computational restrictions, we only considered one of the above species at a time.
At the cloud base pressure ($\Pbase$), we assumed a mass fraction ($\Xbase$) that decreases with altitude following
$ \Xbase \left(\text{P} / \Pbase \right)^{\fsed} $.
Further based on \citet{Ackerman2001PrecipitatingCondensation}, petitRADTRANS computes a distribution of cloud particle sizes based on the sedimentation efficiency $\fsed$, width of the log-normal size distribution $\sigmag$ ($ = 1 + 2 \xsigma$), and vertical eddy diffusion coefficient $\Kzzcloud$.
Although $\Kzz$ is already a fitting parameter in our grid of 1D-RCPE  models, we considered a different eddy diffusion coefficient for the clouds \citep{Zhang2020-Atmosphericregimes}.
The sedimentation efficiency is defined as the ratio of cloud particle sedimentation velocity to turbulent vertical mixing speed. Lower values ($\fsed \lesssim 1$) are indicative of an extended cloud deck, composed of small sub-micron particles while higher values ($\fsed \gtrsim 3-5$) point to a compact cloud deck where larger particles ($\gtrsim 1 \um$) have settled efficiently. 
Initial tests revealed that $\xsigma$ has minimal effect on the final fit. To reduce computation time, we set $\xsigma = -1$.
This totals an additional four free parameters ($\Pbase$, $\Xbase$, $\fsed$, and $\Kzzcloud$) to be retrieved.
\\

\noindent Third, we adopted an alternative cloud treatment, presented by \citet{Welbanks2024-High-InternalHeatFlux}, who parametrize a Gaussian centred around the $10 \um$ silicate cloud feature.
Such a parametrization acknowledges the presence of an additional non-grey opacity source without further specifying its origin.
The parametrization is given by
\begin{equation}
	K_{\rm cloud}(\lambda)= 
	2 \kappaWelbanks \phi \left(\frac{\lambda - \lambda_0}{\omega}\right) \Phi \left(\xi \frac{\lambda - \lambda_0}{\omega}\right)
	\label{eq: Clouds LW}
,\end{equation}
where $\kappaWelbanks$ is the base cloud opacity, $\phi$ the standard normal probability density function, $\Phi$ the standard normal cumulative density function, $\lambda_0$ the mean of the Gaussian, $\omega$ the standard deviation, and $\xi$ a skewness parameter.
After initial testing, and in agreement with \citet{Welbanks2024-High-InternalHeatFlux}, we fixed $\xi = 0$ to reduce the run time of our retrievals with this set-up.
This effectively reduces the parametrization of Eq. \ref{eq: Clouds LW} to a Gaussian function, and 
adds three additional parameters to the fitting routine ($\kappaWelbanks$, $\lambda_0$, and $\omega$).
Henceforth, we refer to this parametrization as `Gaussian` clouds.
\\

\noindent Finally, we parametrized a wavelength-dependent scattering slope by
\begin{equation}
	K_{\rm scatt}(\lambda) = 
	\kappascatt \left(\frac{\lambda}{\lambda_{\rm ref}}\right)^{\gammascatt}
	\label{eq: Scattering parametrization}
,\end{equation}
where $\kappascatt$ is the opacity at $\lambda_{\rm ref} = 0.35 \um$, and $\gammascatt = -4$ in the case for Rayleigh scattering.

\subsection{Data source and model offsets} \label{subsec: Data Sources and model offsets}
We fitted the synthetic transmission spectra of our chemical models to the archival transit spectra of WASP-107b.
Specifically, this includes HST WFC3 ($1.1 - 1.6 \um$) \citep{Kreidberg2018waterw107b}, JWST NIRCam ($2.5 - 5.0 \um$) \citep{Welbanks2024-High-InternalHeatFlux}, and JWST MIRI/LRS ($5.0 - 12.2 \um$) \citep{Dyrek2023so2SilicateClouds}.
For the latter dataset, we used the reduction presented in \citet{Welbanks2024-High-InternalHeatFlux}.
\\

\noindent Aside from the four grid parameters (Z, C/O, $\Tint$, and $\Kzz$), and cloud properties, we also fit for offsets between the HST ($\offsetHST$) and NIRCam ($\offsetNIR$) data with respect to the MIRI data.
The latter is necessary to compensate for differences in instrument sensitivities and assumptions during the data reduction process \citep{Carter2024NatAs...8.1008C}, which can affect the retrieval outcome \citep{Yip_2021}.
Finally, we fitted a reference pressure ($\Pref$) to the planetary radius to account for an offset between the model spectra and data.
We note that some studies choose to fit a planetary radius at a fixed reference pressure, but this choice has no significant effect on the retrieval outcome \citep{Welbanks2019degeneracies}.

\subsection{Nested sampling: UltraNest} \label{subsec: Ultranest}

\noindent To perform the 1D-RCPE retrieval, we used the Python package UltraNest \citep{Buchner2021Ultranest}, which implements the nested sampling algorithm for Bayesian inference.
It estimates the Bayesian evidence (marginal likelihood) and generates approximate posterior distributions of all model parameters.
The algorithm maintains a set of live points that explore the parameter space, iteratively replacing the live point with the lowest likelihood by a new sample with a higher likelihood drawn from the (remaining) prior distributions.
This approach enables UltraNest to navigate complex likelihood surfaces, making it particularly suitable for atmospheric retrieval problems with degenerate and multi-modal posterior distributions.
\\

\noindent
At each likelihood evaluation, a set of parameters is drawn from the prior distributions.
For each combination of Z, C/O, $\Tint$, and $\Kzz$, we quad-linearly interpolate the PT profile, mean-molecular weights, and VMRs for each chemical species from our pre-computed grid of 1D-RCPE models on a common pressure grid (150 log-spaced layers between $10$ and $10^{-7}$ bar) 
before computing the spectrum on-the-fly with petitRADTRANS.
Our default set-up consists of 100 to 200 livepoints (depending on the size of the parameter space), and a convergence criterion $\textsc{dlogz} = 0.8$.
We use high-performance computing facilities and run our retrievals in parallel on a AMD EPYC 7643 3.64GHz CPU, on 80 cores with 250 GB RAM memory or on a Intel Xeon Platinum 8360Y 2.4 GHz CPU, on 130 cores with 232 GB RAM memory.
Run times vary from less than a day to up to two weeks.

\section{Disequilibrium chemistry and pathway analysis} \label{sec: results chemistry}

Before presenting our results of the atmospheric retrievals, we analyse the chemistry simulations of our 1D-RCPE grid of forward models.
More specifically, we discuss the dominant production and destruction pathways of key molecular species that have been detected in WASP-107b.
These include \ce{CH4}, \ce{SO2}, \ce{NH3}, \ce{CO}, \ce{CO2}, and  \ce{H2O} \citep{Kreidberg2018waterw107b, Dyrek2023so2SilicateClouds, Welbanks2024-High-InternalHeatFlux, Sing2024-WarmNeptunes}.
Additionally, we discuss trends with respect to the grid parameters (Z, C/O, $\Tint$, and $\Kzz$) and highlight regions where the composition deviates from chemical equilibrium (i.e. the minimal Gibbs free energy state at local pressure and temperature) by disequilibrium effects such as vertical mixing and photochemistry. 
The goal of this section is to develop a comprehensive understanding of the chemistry in WASP-107b, building on existing literature that has examined a broader range of planetary atmospheres beyond those that resemble WASP-107b.
\\

\begin{figure*}[h!]
	\centering
	\includegraphics[width=17cm]{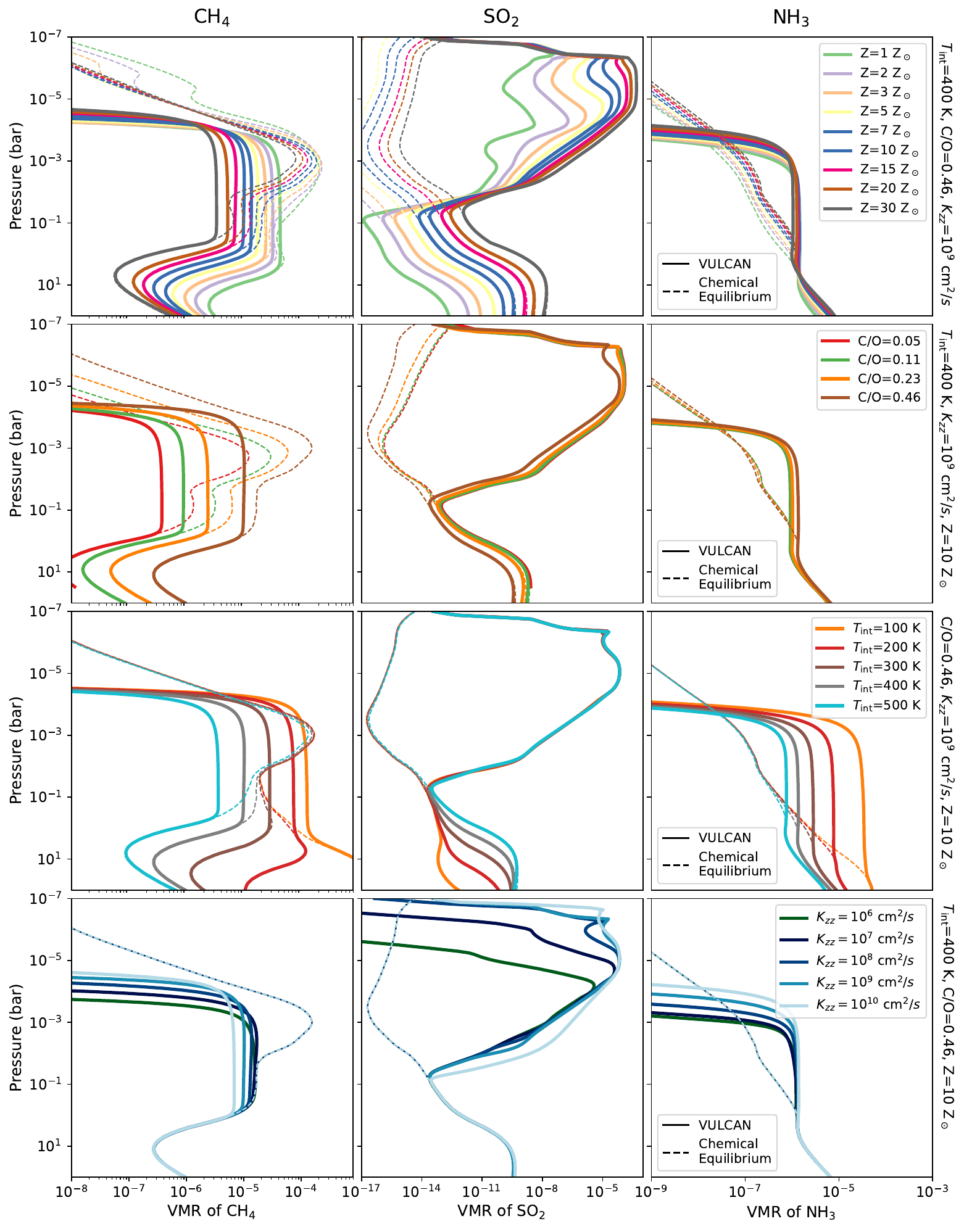}
	\caption{VMR profiles of key species \ce{CH4}, \ce{SO2}, and \ce{NH3} in our grid of disequilibrium chemistry models of WASP-107b. 
		Chemical equilibrium is plotted with thin, dashed lines.
		The nominal model of Z $= 10 \, \Zsun$, C/O = $0.46$, $\Tint=400$ K, and $\log \Kzz = 9 \, (\cmsquareds)$ was chosen to be in agreement with previous studies \citep{Dyrek2023so2SilicateClouds, Welbanks2024-High-InternalHeatFlux}.
	}
	\label{fig: Chem_grid_overview_CH4_SO2_NH3}
\end{figure*}

\noindent Fig. \ref{fig: Chem_grid_overview_CH4_SO2_NH3} shows the behaviour of \ce{SO2}, \ce{CH4}, and \ce{NH3} with respect to the four different grid parameters (Z, C/O, $\Tint$, and $\Kzz$), as these molecules provide information on the sulphur (Sect. \ref{subsec: Sulphur-chemistry}), carbon (Sect. \ref{subsec: Carbon-chemistry}), and nitrogen chemistry (Sect. \ref{subsec: Nitrogen-chemistry}) in WASP-107b.

\subsection{Sulphur chemistry: \ce{H2S}, \ce{SO}, and \ce{SO2}}
\label{subsec: Sulphur-chemistry}
The presence of \ce{SO2} in WASP-107b's atmosphere is evidence of active photochemistry.
Outlined by \cite{Tsai2023PhotochemicallyproducedSO2} for WASP-39b, a photochemical production pathway of \ce{SO2} is
\begin{align} 
	\ce{H2O} &\xrightarrow[]{\text{h}\nu} \ce{OH} + \ce{H} \label{Reaction: Tsai-H2Ophoto}\\
	\ce{H2O} + \ce{H} &\rightarrow \ce{OH} + \ce{H2} \label{Reaction: Tsai-H2O+H->OH+H2} \\
	\ce{H2S} + \ce{H} &\rightarrow \ce{SH} + \ce{H2} \label{Reaction: Tsai-H2S+H->SH+H2}\\
	\ce{SH} + \ce{H} &\rightarrow \ce{S} + \ce{H2} \label{Reaction: Tsai-SH+H->S+H2} \\
	\ce{S} + \ce{OH} &\rightarrow \ce{SO} + \ce{H} \label{Reaction: Tsai-S+OH->SO+H}\\
	\ce{SO} + \ce{OH} &\rightarrow \ce{SO2} + \ce{H} \label{Reaction: Tsai-SO+OH->SO2+H} \\
	\hline
	\text{Net: } \ce{H2S} + 2\ce{H2O} &\rightarrow \ce{SO2} + 3\ce{H2} \label{Reaction: Formation pathway SO2 Tsai2023}
.\end{align}
This shows that \ce{OH}, produced by hydrogen abstraction and photolysis of \ce{H2O}, oxidizes atomic sulphur, which is thermochemically produced from \ce{H2S} via hydrogen abstraction \citep{Hobbs2021-SulfurChemistry, Tsai2023PhotochemicallyproducedSO2}. 
However, for lower temperature atmospheres  (< 1000 K), Re. \ref{Reaction: Tsai-H2O+H->OH+H2} would produce insufficient \ce{OH} to form detectable oxidized S-species \citep{Tsai2023PhotochemicallyproducedSO2}.
The detection of \ce{SO2} in WASP-107b, with $\Teq \approx 740$ K, was enabled by its low gravity (${\sim}\num{270} \cmspersquared$) and favourable stellar UV insolation \citep{Dyrek2023so2SilicateClouds}.
With water photolysis as the dominant source in the upper atmosphere ($ \lesssim \num{e-5}$ bar), \citet{Dyrek2023so2SilicateClouds} argue that the primary source of \ce{OH} between $\num{e-5}$ and $1$ bar 
stems from photolysis of various abundant molecules (including \ce{H2O}, \ce{NH3}, \ce{N2}) aided by Re. \ref{Reaction: Tsai-H2O+H->OH+H2} (see Suppl. Inf. - Figure 14 in \citealt{Dyrek2023so2SilicateClouds}).
A recent study by \citet{deGruyter2024} further investigated the chemical production pathways of \ce{SO2} and, while confirming the two distinct pressure-dependent regimes for \ce{OH} production, revealed an additional complexity regarding the source of atomic sulphur.
Depending on the UV flux and pressure level, sulphur is liberated by hydrogen abstraction of SH (Re. \ref{Reaction: Tsai-SH+H->S+H2}) and/or by photolysis of \ce{SH} and \ce{S2} that liberate sulphur from the \ce{H2S} reservoir.
\\

\noindent From Fig. \ref{fig: Chem_grid_overview_CH4_SO2_NH3}, it is clear that \ce{SO2} is most sensitive to the atmospheric metallicity.
The VMRs show a positive correlation with metallicity that primarily results from the larger sulphur reservoir (\ce{H2S}), but also from the increase in hydroxide (\ce{OH}), in line with previous studies \citep{Polman2023-H2SandSO2detectabilityinHJs, Dyrek2023so2SilicateClouds}.
We validate the formation pathway of \ce{SO2} along this dimension of our grid, using models with C/O = $0.46$, $\Tint = 400$ K, and $\log \Kzz = 9$ (cgs), similar parameters as presented in \citet{Dyrek2023so2SilicateClouds}.
We find that \ce{OH} radicals are produced by hydrogen abstraction of \ce{H2O} (Re. \ref{Reaction: Tsai-H2O+H->OH+H2}) throughout the atmosphere and direct photolysis of \ce{H2O} (Re. \ref{Reaction: Tsai-H2Ophoto}), where the latter dominates the upper layers.
At solar metallicity (Z = $1 \, \Zsun$), the region where Re. \ref{Reaction: Tsai-H2Ophoto} dominates extends up to several millibar. 
For extremely metal-rich models ($Z > 10 \, \Zsun$), this region is limited to only the very upper layers ($\lesssim \mubar$), while Re. \ref{Reaction: Tsai-H2O+H->OH+H2} dominates \ce{OH} production elsewhere.
This indicates that radiation penetrates to shallower pressures in higher-metallicity atmospheres, which results from the additional metal opacity that increases the overall optical depth at any given pressure layer.
Hence, for higher atmospheric metallicities, the layer at which the atmosphere becomes optically thick moves upward. 
\\

\noindent We find that the production pathway from \ce{OH} to \ce{SO2} depends on metallicity and pressure in our models.
At solar metallicity (Z = 1 $\Zsun$), the entire atmospheric column follows the production pathway summarized by Re. \ref{Reaction: Formation pathway SO2 Tsai2023}.
Given the quadratic dependence of sulphur atoms on metallicity, we see an increased reaction rate for reactions with exclusively sulphur reactants for higher Z models (> 7 $\Zsun$).
In deeper layers of such high-metallicity atmospheres, \ce{SO} is formed predominantly by 
\begin{align}
	\ce{S} + \ce{SH} &\rightarrow \ce{H} + \ce{S2}   \label{Reaction:S+SH->H+S2} \\
	\ce{S2} + \ce{OH} &\rightarrow \ce{SO} + \ce{SH} \label{Reaction:S2+OH->SO+SH}
,\end{align}
where S arises predominantly from Re. \ref{Reaction: Tsai-SH+H->S+H2} or \ce{SH} recombination
\begin{align}
	\ce{SH} + \ce{SH} \rightarrow \ce{H2S} + \ce{S} \text{.} \label{Reaction:SH+SH->H2S+S}
\end{align}
Finally, in high-metallicity environments, the subsequent production of \ce{SO2} results primarily from
\begin{align}
	\ce{SO} + \ce{SO}\rightarrow \ce{SO2} + \ce{S}  \label{Reaction:SO+SO->SO2+S}
\end{align}
rather than Re. \ref{Reaction: Tsai-SO+OH->SO2+H}.
\newline

\begin{figure*}[h!]
	\centering
	\includegraphics[width=17cm]{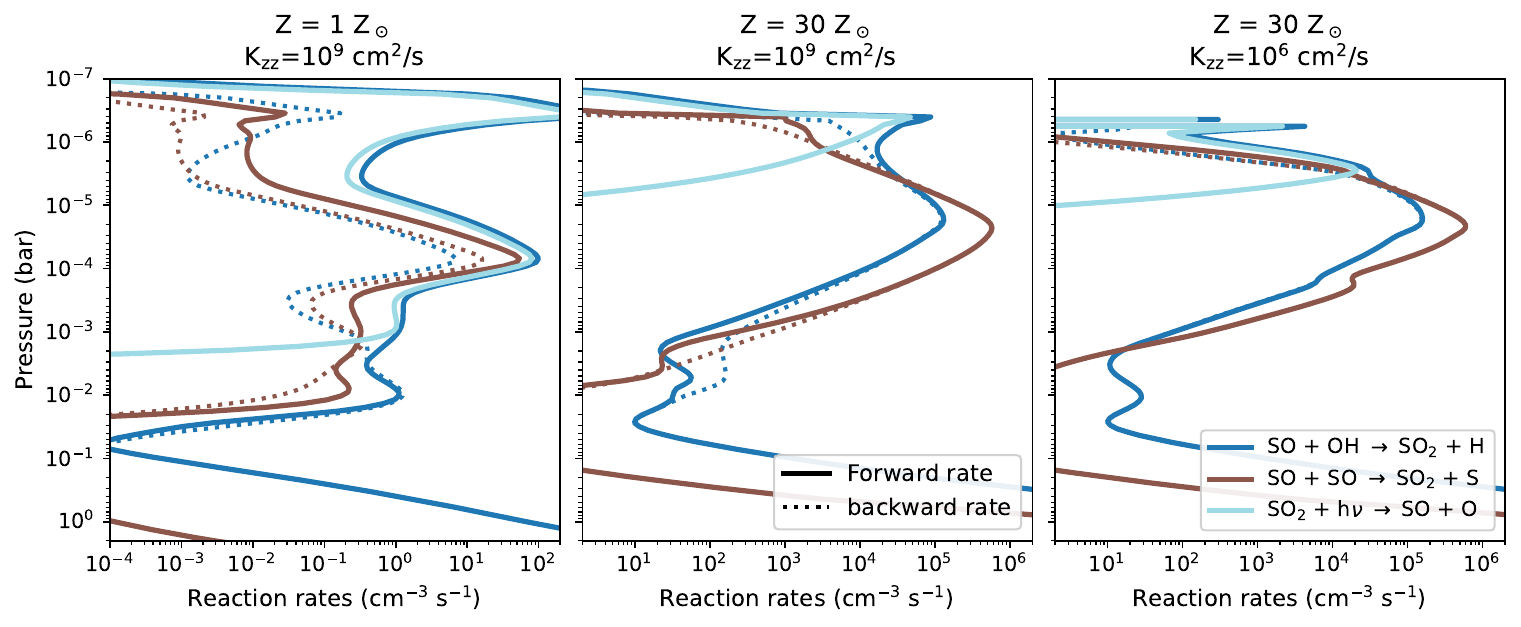}
	\caption{
		Dominant reactions (by fastest reaction rate) that involve \ce{SO} and \ce{SO2}, for three different combinations of metallicity and eddy diffusion coefficient.
		Forward rates are shown by solid lines; backward rates by dotted lines.
		The other values of these simulations are $\Tint = 400$ K and C/O = 0.46.
	}
	\label{fig: Chemistry_grid_SO2producingreactions}
\end{figure*}
\begin{figure*}[h!]
	\centering
	\includegraphics[width=17cm]{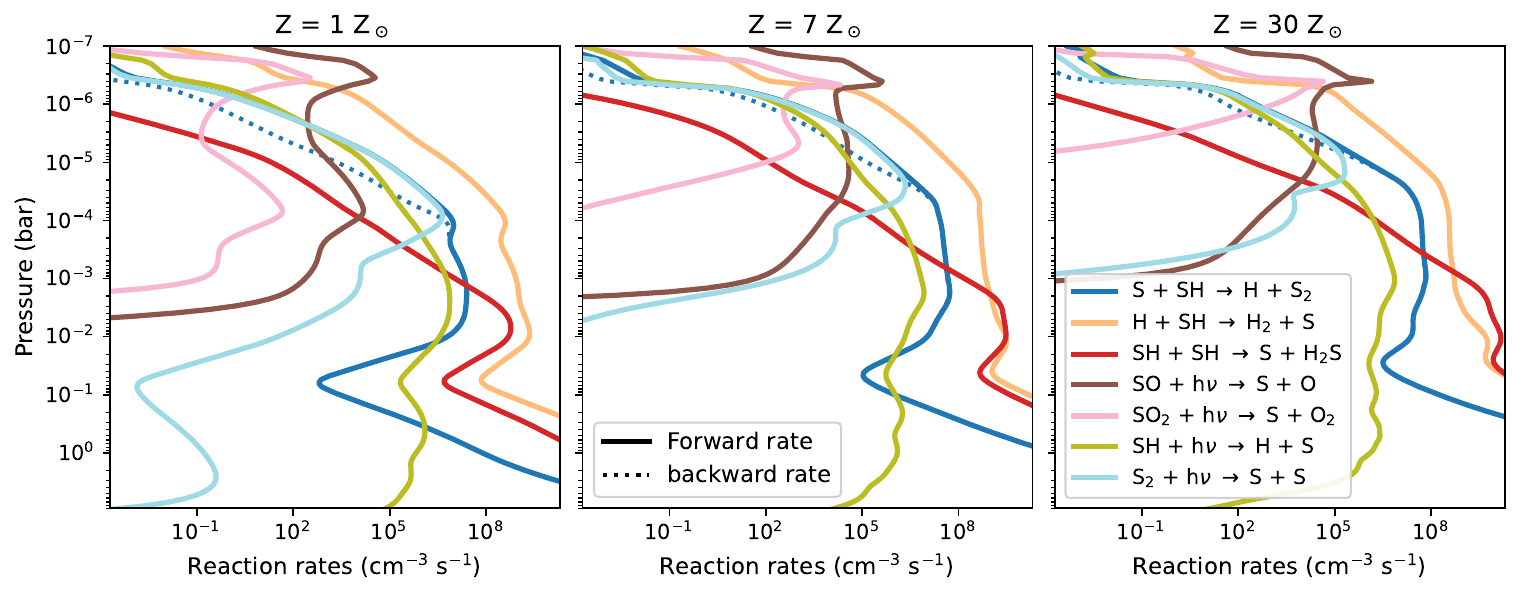}
	\caption{
		Dominant reactions (by fastest reaction rate) that produce/destroy S, for three different metallicities.
		Forward rates are shown by solid lines; backward rates by dotted lines.
		The other values of the simulations are $\Tint = 400$  K, C/O = 0.46 and $\log \Kzz = 9 \, (\cmsquareds)$.
	}
	\label{fig: Chemistry_grid_Sproducingreactions}
\end{figure*}

\noindent
One can further analyse the last step of the production pathway ($\ce{SO} \rightarrow \ce{SO2}$) by looking at the reaction rates in Fig. \ref{fig: Chemistry_grid_SO2producingreactions}, which shows the fastest reactions (by reaction rate) involving \ce{SO2} and \ce{SO}.
At solar metallicity (left panel, Fig. \ref{fig: Chemistry_grid_SO2producingreactions}), Re. \ref{Reaction: Tsai-SO+OH->SO2+H} (forward) is faster than Re. \ref{Reaction:SO+SO->SO2+S} (forward), making it the dominant \ce{SO2}-producing reaction.
By dominant, we mean that removing this particular reaction from the network will significantly disturb the steady-state equilibrium abundances.
Furthermore, both reactions run much slower in their reverse directions.
This indicates that, when isolating these reactions, a net production of \ce{SO2} is achieved.
To reach a steady state, and constant VMR, direct photolysis ($\text{S}\text{O}_2 \xrightarrow[]{\text{h}\nu} \text{SO} + \text{O}$) is the main sink of \ce{SO2} and balances the system almost entirely.
\newline

\noindent For Z = $30 \, \Zsun$ (middle panel, Fig. \ref{fig: Chemistry_grid_SO2producingreactions}), between ${\sim}\mbar$ and ${\sim}100 \mubar$, the opposite situation occurs where Re. \ref{Reaction:SO+SO->SO2+S} is slightly faster (in both directions) than Re. \ref{Reaction: Tsai-SO+OH->SO2+H}, making it the dominant reaction that controls the VMR of \ce{SO2}.
This statement is valid for both production and destruction of \ce{SO2}, as Re. \ref{Reaction:SO+SO->SO2+S} is nearly perfectly balanced in both directions as a result of little to no direct photolysis in these regions.
This reasoning does not hold for a small region between $\num{e-2}$ and $\num{e-3}$ bar, where it appears that both Re. \ref{Reaction: Tsai-SO+OH->SO2+H} and \ref{Reaction:SO+SO->SO2+S} are destroying \ce{SO2} at a faster reaction rate than producing it.
It is also evident that no other thermo- or photochemical reactions can compensate this net destruction of \ce{SO2}, as we have plotted the fastest reaction at play.
An explanation for the imbalance in \ce{SO2} production and destruction is found in the disequilibrium effect of vertical mixing.
Indeed, when comparing to simulations with lower $\Kzz$ (right panel, Fig. \ref{fig: Chemistry_grid_SO2producingreactions}) where the atmosphere is not quenched in this particular model, we see that both reactions are perfectly balanced in both directions.
For quenched atmospheres, such as with $\log \Kzz = 9 \, (\cmsquareds)$, we see the immediate effect of \ce{SO2} replenishment due to vertical upward mixing.
We conclude that the dominant \ce{SO2} production pathway is influenced by atmospheric metallicity and that vertical mixing plays a crucial role in transporting \ce{SO2} upward.
\newline

\noindent
One can also further analyse the species whose sulphur ends up in \ce{SO} (and later \ce{SO2}).
Both pathways via Re. \ref{Reaction: Tsai-S+OH->SO+H} as well as Re. \ref{Reaction:S+SH->H+S2} and \ref{Reaction:S2+OH->SO+SH}, require atomic sulphur whose origin is debated in literature \citep{Tsai2023PhotochemicallyproducedSO2, deGruyter2024}.
Fig. \ref{fig: Chemistry_grid_Sproducingreactions} shows that we find Re. \ref{Reaction: Tsai-SH+H->S+H2} to be dominant throughout the atmosphere with low metallicity (e.g. Z = $1 \, \Zsun$), while Re. \ref{Reaction:SH+SH->H2S+S} becomes faster at higher metallicity for pressures above ${\sim} \, 1 \mbar$.
Furthermore, both the forward and backward reaction rates of Re. \ref{Reaction: Tsai-SH+H->S+H2} are comparable, thus acting as the primary source and sink of \ce{SH} in the regions where the VMR of \ce{SO2} peaks (see Fig. \ref{fig: Chem_grid_overview_CH4_SO2_NH3}).
Further analysis of Re. \ref{Reaction: Tsai-SH+H->S+H2} reveals that it is every so slightly out of equilibrium (< 2 \%) favouring the backwards direction.
However, we believe this is not an argument against the ability of Re. \ref{Reaction: Tsai-SH+H->S+H2} to liberate S from \ce{SH} \citep{deGruyter2024}. 
Instead, this indicates that additional S-producing reactions (of secondary importance) compensate for this slight imbalance to reach a steady-state equilibrium.
Finally, our models show that hydrogen abstraction of \ce{H2S} (Re. \ref{Reaction: Tsai-H2S+H->SH+H2}) is the fastest way to produce \ce{SH} from the \ce{H2S} reservoir.
The different chemical pathways that we have identified are summarized in Table \ref{tab: metallicityDimension: how to form-SO2}.
\newline

\begin{table}[t]
	\begin{center}
		\centering
		\caption{\label{tab: metallicityDimension: how to form-SO2} Dominant reactions that convert \ce{OH} into \ce{SO2} for different metallicities and pressure levels. Other parameters of these nominal models are C/O = 0.46, $\Tint = 400$ K, and $\log \Kzz = 9$ (cgs).
		}
		\begin{tabular}{lcc}
			\hline\hline\rule[0mm]{0mm}{5mm}
			Z (Z$_\odot$) & $\num{e-3}$ bar & $\num{e-5}$ bar \\
			\hline\hline\rule[0mm]{0mm}{5mm}
			\noindent
			1, 2 & $\ce{OH} + \ce{S} \rightarrow \ce{H}+ \ce{SO}$ & (Same to $\num{e-3}$ bar) \\
			& $\ce{SO} + \ce{OH} \rightarrow \ce{SO2} + \ce{H}$ &  \\
			&& \\
				
			\hline\rule[0mm]{0mm}{5mm}
			3, 5 & $\ce{S} + \ce{SH} \rightarrow \ce{H}+ \ce{S2}$  &  $\ce{OH} + \ce{S} \rightarrow \ce{H}+ \ce{SO}$  \\
			& $\ce{S2} + \ce{OH} \rightarrow \ce{SO} + \ce{SH}$  &  $\ce{SO} + \ce{OH} \rightarrow \ce{SO2} + \ce{H}$   \\
			&  $\ce{SO} + \ce{SO} \rightarrow \ce{SO2} + \ce{S}$  &  \\
			&& \\
			
			\hline\rule[0mm]{0mm}{5mm}
			$\ge 7$ & $\ce{S} + \ce{SH} \rightarrow \ce{H}+ \ce{S2}$   &  $\ce{OH} + \ce{S} \rightarrow \ce{H}+ \ce{SO}$ \\
			& $\ce{S2} + \ce{OH} \rightarrow \ce{SO} + \ce{SH}$ & $\ce{SO} + \ce{SO} \rightarrow \ce{SO2} + \ce{S}$ \\
			&  $\ce{SO} + \ce{SO} \rightarrow \ce{SO2} + \ce{S}$    &  \\
			&& \\
			
			\hline\rule[0mm]{0mm}{3mm}
		\end{tabular}
	\end{center}
\end{table}

\begin{figure}[h!]
	\resizebox{\hsize}{!}{\includegraphics{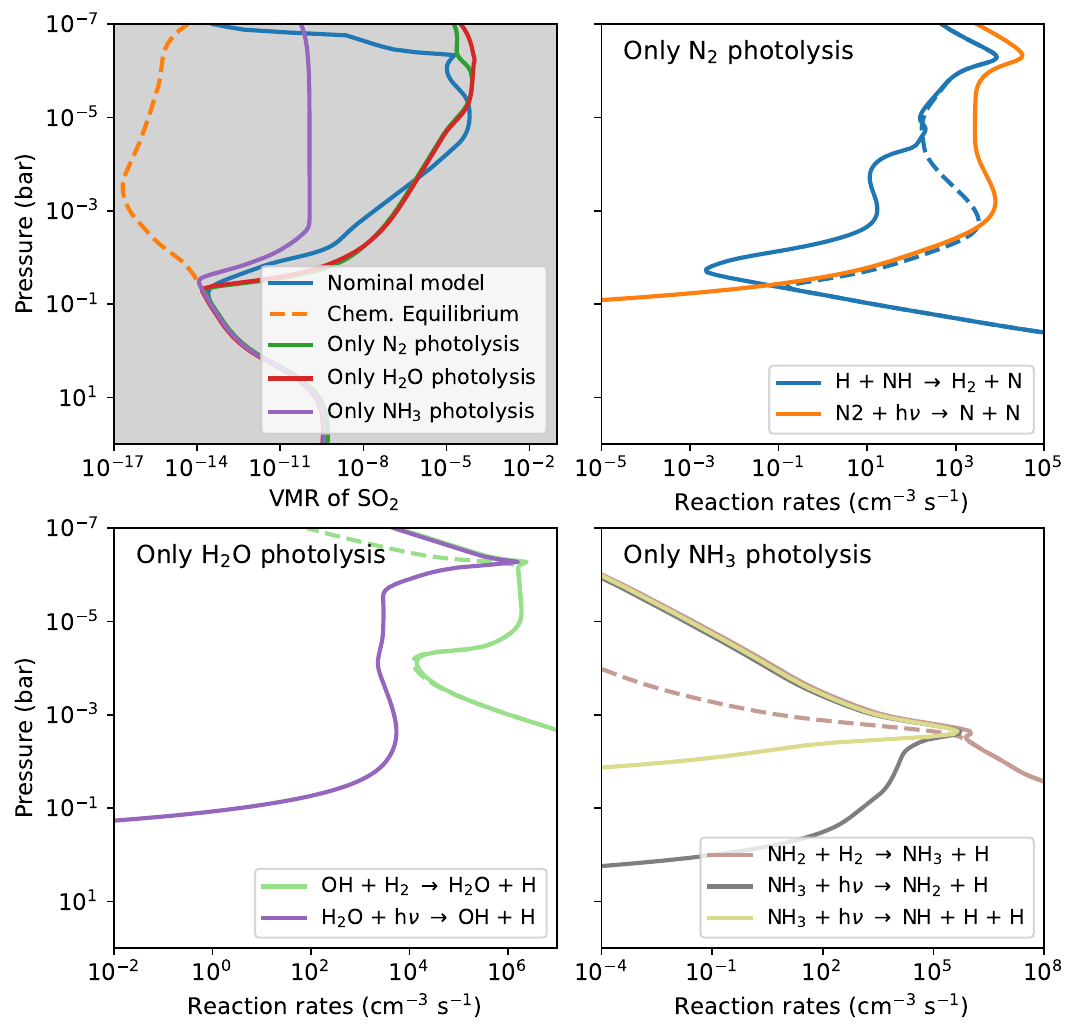}}
	\caption{VMR of \ce{SO2} and dominant H-producing reactions for a nominal model (Z $= 10 \, \Zsun$, C/O $=0.46$, $\Tint=400$ K, $\log \Kzz = 9 \cmsquareds$) that includes all available photodissociation cross-sections (upper left panel), and three limiting cases where only the photolysis of one particular source is considered (\ce{N2}, \ce{H2O}, or \ce{NH3}), copying the experiment of \citet{Dyrek2023so2SilicateClouds}.
	Panels that show thermochemical reaction rates show both their forward (solid lines) and backward (dashed) rates.
	}
	\label{fig: Chemistry_grid_cascade_scenario}
\end{figure}

\noindent
This paints a complete picture of the dominant chemical pathways that lead to \ce{SO2} in our models of WASP-107b.
One final remark can be made regarding Re. \ref{Reaction: Tsai-H2O+H->OH+H2}, which converts \ce{H2O} into \ce{OH} by hydrogen abstraction.
It is clear that an excess of H is available throughout the atmosphere, whose origin is photochemical.
However, \citet{Dyrek2023so2SilicateClouds} revealed the somewhat puzzling result that models with photochemistry of non-hydrogen bearing species can still produce significant amounts of \ce{SO2}.
For example, the authors show (Suppl. Inf. - Figure 14 in \citealt{Dyrek2023so2SilicateClouds}) that only \ce{N2}-photolysis is sufficient to produce \ce{SO2}, but do not specify the chemical pathway.
This calls for further investigation on how sufficient amounts of H, and subsequently \ce{OH} via Re. \ref{Reaction: Tsai-H2O+H->OH+H2}, can be produced in the \ce{SO2}-producing layers from photolysis of non-hydrogen bearing species.
In Fig. \ref{fig: Chemistry_grid_cascade_scenario}, we use our model set-up
to run three models where we consider only \ce{N2}, \ce{H2O}, or \ce{NH3} photochemistry.
When considering only \ce{N2} as a photon absorber, we validate that the products of \ce{N2} photolysis can enable \ce{SO2} production, with the help of the subsequent thermochemical reaction,
\begin{align}
	\ce{H2}+\ce{N} \rightarrow \ce{H} + \ce{NH} \label{Reaction: H2+N->H+NH}
,\end{align}
which acts as the dominant H-producing reaction by interacting with the abundant \ce{H2} reservoir.
We do note, however, that there are several faster H-producing reactions when the final steady state is reached in our simulations. 
However, we are specifically interested in pathways that start from species that are abundant in chemical equilibrium, from which the chemistry simulation is initialized, such as $\ce{H2}$.
\newline

\noindent When we only include \ce{H2O} photolysis (Re. \ref{Reaction: Tsai-H2Ophoto}), Fig. \ref{fig: Chemistry_grid_cascade_scenario} shows that both \ce{OH} and H are deposited deep in the atmosphere, which activates Re. \ref{Reaction: Tsai-H2O+H->OH+H2} in both directions.
In models with similar parameters where we include all available photolysis reactions (Sect. \ref{subsec: Sulphur-chemistry}), stellar radiation is absorbed much higher up in the atmosphere.
When we only allow for \ce{H2O} photolysis, radiation can penetrate much deeper due to the absence of other photon absorbers in this specific model.
\newline

\noindent Compared to \citet{Dyrek2023so2SilicateClouds}, we produce much less \ce{SO2} when only activating \ce{NH3} photolysis.
However, our network analysis reveals that radiation penetrates deep in the atmosphere, releasing H there via both of the following reactions:
\begin{align}
	\ce{NH3} &\xrightarrow[]{\text{h}\nu} \ce{NH2} + \ce{H} \label{Reaction: NH3 -> NH2 +  H} \\
	\ce{NH3} &\xrightarrow[]{\text{h}\nu} \ce{NH} + \ce{H} + \ce{H}  \label{Reaction: NH3 -> NH +  H + H} \, \text{.}
\end{align}

\noindent
Once again, we note that such deep penetration depth can only be reached in absence of other photon absorbers.
The lack of H deposition in the upper layers, and resulting lack of \ce{SO2} with respect to \citet{Dyrek2023so2SilicateClouds} can be explained due to the absence of sub-100 nm photoabsorption cross-sections in our chemical network.
We also note that vertical mixing cannot efficiently transport H atoms upwards, due to their short chemical timescales, but can successfully quench \ce{SO2} in these layers. 
Additionally, the photochemical product $\ce{NH2}$ can locally produce H atoms by $\ce{NH2} +\text{H}_2 \rightarrow \ce{NH3} + \text{H}$.
In reality, a mixture of \ce{N2}, \ce{H2O}, \ce{NH3} and other abundant photochemically active species all contribute to liberating H atoms throughout the atmosphere although Re. \ref{Reaction: Tsai-H2O+H->OH+H2} remains the dominant source.

\subsection{Carbon chemistry: \ce{C2H2}, \ce{CO}, \ce{CO2}, and \ce{CH4}}
\label{subsec: Carbon-chemistry}
From Fig. \ref{fig: Chem_grid_overview_CH4_SO2_NH3}, it is clear that all four dimensions of the parameter space (Z, C/O, $\Tint$, $\Kzz$) impact the VMR profile of \ce{CH4}.
Furthermore, each panel shows the three well-known regimes from the top to bottom layers: photochemical destruction, transport-induced quenching, and thermochemical equilibrium.
Aside from the general carbon budget, controlled by metallicity and carbon-to-oxygen ratio, the VMR of \ce{CH4} is mainly controlled by the VMR around the quench pressure, i.e. the point at which the chemical timescale matches the vertical mixing timescale.
\\

\begin{figure}[h]
	\resizebox{\hsize}{!}{\includegraphics{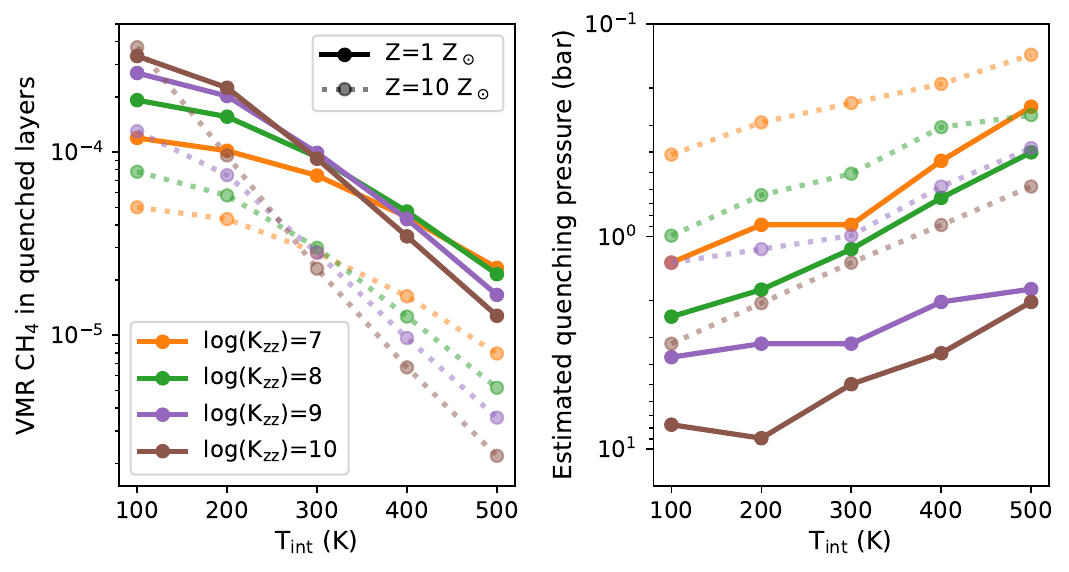}}
	\caption{Quenching VMR (left panel) and estimated quenching pressure (right panel) of \ce{CH4} versus intrinsic temperature $\Tint$ for various eddy diffusion coefficients in low (Z = 1 Z$_\odot$) and high (Z = 10 Z$_\odot$) metallicity atmospheres. The models with $\log \Kzz =6$ (cgs) are not included as not all atmospheres are quenched at such weak vertical mixing.}
	\label{fig: CH4_Quenching}
\end{figure}

\noindent The absence of abundant \ce{CH4} in WASP-107b has led to the suggestion of a tidally heated core (high $\Tint$), in combination with strong vertical mixing (high $\Kzz$) that quenches \ce{CH4} deep in the atmosphere and thus reduces the overall \ce{CH4} content in the observable layers \citep{Fortney2020BeyondEq, Dyrek2023so2SilicateClouds, Welbanks2024-High-InternalHeatFlux, Sing2024-WarmNeptunes}.
This is also apparent from Fig. \ref{fig: CH4_Quenching}, which shows the quenched VMR and pressures of \ce{CH4} in our chemistry grid.
At high $\Tint$ (> 300 K), faster vertical mixing indeed quenches \ce{CH4} more deeply (higher pressure), resulting in an overall lower VMR.
The opposite is true for low $\Tint$ ($\le 200$ K), where we see that deeper quenching results in more \ce{CH4}.
This is due to a negative VMR gradient for layers in chemical equilibrium, as compared the positive gradient in models with high $\Tint$ (see also Fig. \ref{fig: Chem_grid_overview_CH4_SO2_NH3}).
\\

\noindent Furthermore, Fig. \ref{fig: CH4_Quenching} shows an upwards shift in altitude (downwards in pressure) of the quench point at higher metallicity, together with an overall lower quenched VMR of \ce{CH4}.
The negative correlation between metallicity and \ce{CH4} is counter-intuitive as one expects a positive correlation between the available carbon budget and eventual VMR of \ce{CH4}.
However, this is explained by Fig. \ref{fig: temperatures_grid}, which shows that the deep layers are warmer at higher Z as a result by of the increased abundance (and thus opacity) of strong absorbers \citep{Drummond2018EffectofMetallicity}.
Additionally, chemistry in high-metallicity environments shifts the \ce{CH4}-\ce{CO} boundary to lower temperatures \citep{Lodders2002Icar..155..393L}.
A warmer interior, for similar eddy diffusion coefficients, and high metallicity explain the upwards shift in altitude of the quench point, resulting in less \ce{CH4}.
\newline

\noindent
The formation of \ce{CH4} is coupled to \ce{CO}, as both molecules can act as the dominant carbon reservoir under different circumstances.
We find that in the majority of our models, \ce{CO} dominates over \ce{CH4}.
Only in set-ups with (i) quasi-solar metallicity and/or (ii) shallow quenching by low $\Tint$ or weak $\Kzz$, is \ce{CH4} more abundant than \ce{CO}.
For our nominal model of Z = 10 $\Zsun$, C/O = 0.46, $\Tint = 400$ K, and $\log \Kzz =9$ (cgs), we analyse the conversion pathway from \ce{CH4} to \ce{CO} at the quench pressure and find 
\begin{align}
	\ce{CH4} + \ce{H}&\rightarrow \ce{CH3}+ \ce{H2}  \label{Reaction:CH4+H->CH3+H2} \\
	\ce{CH3}+ \ce{OH} &\rightarrow \ce{CH2OH} + \ce{H} \label{Reaction:CH3+OH->CH2OH+H} \\
	\ce{CH2OH} + \text{M} &\rightarrow \ce{H}+ \ce{H2CO} + \text{M} \label{Reaction:CH2OH+M->H+H2CO+H} \\
	\ce{H2CO} + \ce{H}&\rightarrow \ce{HCO} + \ce{H2} \label{Reaction:H2CO+H->HCO+H2} \\
	\ce{HCO} + \text{M} &\rightarrow \ce{H}+ \ce{CO}+ \text{M} \label{Reaction:HCO+M->H+CO+M} \\
	\hline
	\text{Net: } \ce{CH4} + \ce{H2O} &\rightarrow \ce{CO}+ 3\ce{H2} \label{Reaction: ConversionCH4toCOwithoutS}
,\end{align}
taking into account that \ce{OH} is produced by Re. \ref{Reaction: Tsai-H2O+H->OH+H2}.
This pathway has been well documented in previous studies \citep{Moses2011-Disequilibrium, Venot2012-Achemicalmodel, Zahnle2014-MethaneCarbonMonoxid, Tsaj2017-VULCANaopensource, Tsai2018-TowardConsistentModeling, Venot2020-Newchemicalscheme}. 
Some of the above papers have reported dominant pathways that run via \ce{CH3OH}, effectively replacing Re. \ref{Reaction:CH3+OH->CH2OH+H} with 
\begin{align}
	\ce{CH3}+ \ce{OH} + \text{M} &\rightarrow \ce{CH3OH} + \text{M}  \label{Reaction:CH3+OH+M->CH3OH+M} \\
	\ce{CH3OH} + \ce{H}&\rightarrow \ce{CH2OH} + \ce{H2}  \label{Reaction:CH3OH+H->CH2OH+H2} 
\end{align}
for high-pressure (deep quenching) environments.
We do not find a model in which this pathway is faster than Re. \ref{Reaction:CH3+OH->CH2OH+H}.
A faster pathway exists via COS, a stable sulphur molecule present in low amounts at chemical equilibrium 
(VMR ${<}\num{e-7}$ in our nominal model), 
which produces \ce{CO} via $\ce{COS} + \ce{CH3}\rightarrow \ce{CO}+ \ce{CH3S}$.
However, COS itself is predominantly kept in equilibrium by reactions $\ce{H}+ \ce{COS}\leftrightarrow \ce{CO}+ \ce{SH}$ and $\ce{COS}+\ce{SH} \leftrightarrow \ce{CO}+ \ce{HS2}$, which both have \ce{CO} as a reactant, disqualifying COS from the interconversion pathway \ce{CH4}-\ce{CO}.
Finally, we find that multiple other pathways can come close to competing with the pathway of Re. \ref{Reaction: ConversionCH4toCOwithoutS}.
This includes, for example, stripping \ce{CH4} to atomic carbon before oxidization to \ce{CO}.
We refer to \citet{Tsai2018-TowardConsistentModeling} and \citet{Hu2021PhotochemistryAndSpectralCharacterization} for a detailed description on the various \ce{CH4}-\ce{CO} pathways and regimes in which they occur.
\newline

\noindent
Other carbon-species of interest are \ce{CO2} (detected in WASP-107b) and $\ce{C2H2}$ (undetected).
Production and destruction of \ce{CO2} is equilibrated by 
\begin{align}
	\ce{OH} + \ce{CO}\rightarrow \ce{CO2} + \ce{H} \label{Reaction:OH+CO->CO2+H}
\end{align}
throughout the entire atmosphere across all dimensions of our model grid, and its VMR is hardly altered by disequilibrium chemistry (see Fig. \ref{appendixfig: Chemistry_grid_plot_HCN_CO2_C2H2_SO_H2S}).
However, we note photochemically produced \ce{OH} has been shown to increase \ce{CO2} in some exoplanetary atmospheres \citep{Moses2011-Disequilibrium, Zahnle2014-MethaneCarbonMonoxid, Hu2021PhotochemistryAndSpectralCharacterization}. 
Finally, we briefly discuss \ce{C2H2} as it another important photochemical product in exoplanetary atmospheres \citep{Line2010HighTemperaturePhotochemistry, Moses2014-Chemicalkinetics, Baeyens2022-GridIIphotochemistry, Konings2022ImpactofStellarFlares}.
We find the average VMR of \ce{C2H2} to be below $\num{e-10}$ in all our models, making it too scarce to be observable.
We do note that its primary formation pathway in our models is well known, following $\ce{CH4} \rightarrow \ce{CH3}\rightarrow \ce{C2H6} \rightarrow \ce{C2H5} \rightarrow \ce{C2H4} \rightarrow \ce{C2H3} \rightarrow \ce{C2H2}$, as was described in  \citet{Moses2011-Disequilibrium}.

\subsection{Nitrogen chemistry: \ce{HCN}, \ce{N2} and \ce{NH3}}
\label{subsec: Nitrogen-chemistry}

\citet{Dyrek2023so2SilicateClouds} reported a tentative detection of \ce{NH3} ($2-3 \sigma$) in WASP-107b at a VMR of $\lesssim 10^{-5.47}$, based on the $10.5 \um$ feature in the MIRI data.
Subsequently, \citet{Sing2024-WarmNeptunes} found marginal evidence of \ce{NH3}, based on a $3 \um$ feature in the NIRSpec data.
Most recently, \citet{Welbanks2024-High-InternalHeatFlux} could confirm the presence of \ce{NH3} with a $6 \sigma$ detection based on a panchromatic analysis covering both the $3$ and $10.5 \um$ feature at $\lesssim \num{e-5}$ VMR, while also requiring an enhancement of a factor ${\sim} 30$ to match the observed VMR to their best-fit forward model.
This sparks an interest in the behaviour and chemical pathway of \ce{NH3} in our models.
\\

\noindent Returning to Fig. \ref{fig: Chem_grid_overview_CH4_SO2_NH3}, we observe only a slight dependency of \ce{NH3} on metallicity despite the linear dependence with the available nitrogen budget.
Similarly to the behaviour of \ce{CH4}, outlined in Sect. \ref{subsec: Carbon-chemistry}, one must take into account the PT profile and its vertical upward shift for higher metallicity as well as the shift in \ce{NH3}-\ce{N2} balance \citep{Lodders2002Icar..155..393L}.
Given the quasi-constant VMRs in the layers around the \ce{NH3} quench point, the upward (or downward) shift of the quench point altitude (or pressure) does not substantially affect the quenched VMR.
A similar explanation can be given to the behaviour with respect to $\Kzz$ as a stronger eddy diffusion coefficient, and thus deeper quenching, does not affect the quenched \ce{NH3} VMR too much due to quasi-constant VMRs in chemical equilibrium \citep{Moses2011-Disequilibrium, Moses2014-Chemicalkinetics}.
\\

\begin{figure}[h]
	\resizebox{\hsize}{!}{\includegraphics{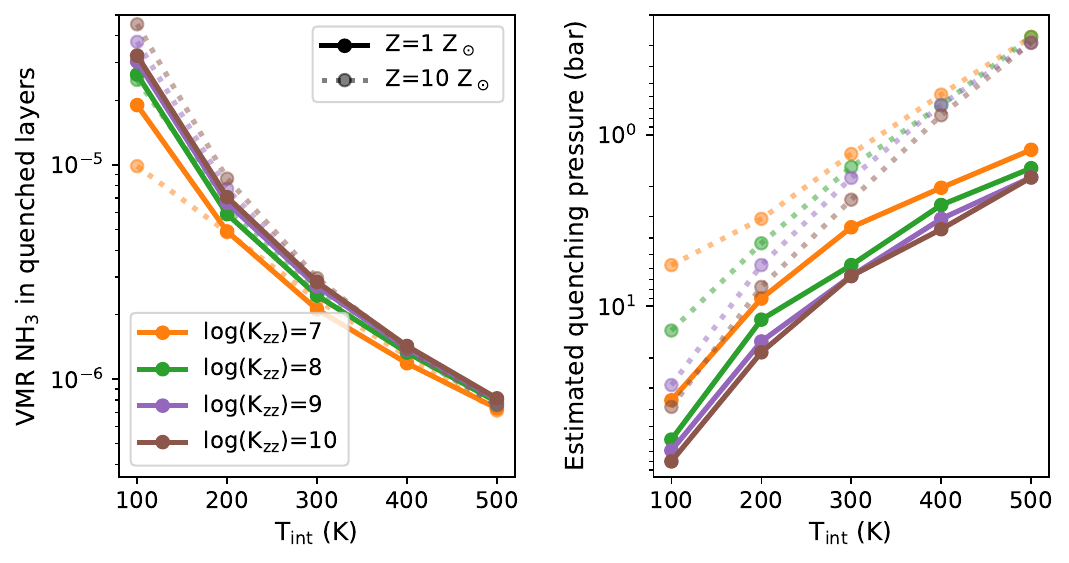}}
	\caption{Same as Fig. \ref{fig: CH4_Quenching}, now for \ce{NH3}.}
	\label{fig: NH3_Quenching}
\end{figure}

\noindent Similar to the \ce{CH4}-\ce{CO} interconversion described in Sect. \ref{subsec: Carbon-chemistry}, the ratio between \ce{NH3} and \ce{N2} is frozen at the quench point and sets the dominant nitrogen reservoir in the middle atmosphere.
From Fig. \ref{fig: NH3_Quenching}, it is clear that a higher intrinsic temperature results in a lower quenched VMR of $\ce{NH3}$ as the quench point moves up in altitude.
Additionally, the minimal dependence on metallicity and $\Kzz$ is clearly shown here as well.

\noindent 
In all of our chemistry models, we find a clear overabundance of \ce{N2} with respect to \ce{NH3} by several orders of magnitude.
Only in a limiting case with Z = $1 \, \Zsun$, $\Tint=100$ K, and high $\Kzz$, the overabundance of \ce{N2} is limited to a factor two.
We identify the primary pathway for \ce{NH3}-\ce{N2} conversion as
\begin{align}
	2 \times \left( \ce{NH3} + \ce{H}\right. &\rightarrow \left. \ce{NH2} + \ce{H2} \right)  \label{Reaction:NH3+H->NH2+H2} \\
	\ce{H}+ \ce{NH2} &\rightarrow  \ce{H2} + \ce{NH}      \label{Reaction:H+NH2->H2+NH}  \\
	\ce{NH} + \ce{NH2} &\rightarrow \ce{N2H2} + \ce{H}  \label{Reaction:NH+NH2->N2H2+H} \\
	\ce{N2H2} + \ce{H}&\rightarrow \ce{N2H} + \ce{H2} \label{Reaction:N2H2+H->N2H+H2} \\
	\ce{N2H} + \text{M} &\rightarrow \ce{N2} + \ce{H}+ \text{M} \label{Reaction:N2H+M->N2+H+M} \\
	\hline
	\text{Net: } 2 \ce{NH3} &\rightarrow \ce{N2} + 3\ce{H2} \label{Reaction: ConversionNH3toN2}
\end{align}
around the quench pressure, which is in agreement with the literature \citep{Moses2011-Disequilibrium, Hu2021PhotochemistryAndSpectralCharacterization}.
An alternative pathway can replace Re. \ref{Reaction:NH+NH2->N2H2+H} with 
\begin{align}
	\ce{NH2} + \ce{NH3} &\rightarrow \ce{N2H3} + \ce{H2} \label{Reaction:NH2+NH3->N2H3+H2} \\
	\ce{N2H3} + \text{M} &\rightarrow \ce{N2H2} + \text{H}+\text{M} \text{,} \label{Reaction:N2H3+M->N2H2+H+M} 
\end{align}
at higher quenching pressures but fails to be the quickest way to form \ce{N2} from \ce{NH3} in our models.
\newline

\noindent
A final nitrogen-species of potential interest is $\ce{HCN}$, as photochemistry can enhance its presence significantly from chemical equilibrium \citep{Baeyens2024Photodissociation}.
However, \ce{HCN} remains undetected in WASP-107b, and is not expected to be prominent in transit spectra at this temperature regime \citep{Baeyens2022-GridIIphotochemistry}.
In our models, \ce{HCN} reaches an maximal VMR of ${\sim}\num{e-6}$ in some limiting cases (see Fig. \ref{appendixfig: Chemistry_grid_plot_HCN_CO2_C2H2_SO_H2S}).
It forms primarily from the net reaction $\ce{CH4} + \ce{NH3} \rightarrow \ce{HCN} + 3 \ce{H2}$, after hydrogen abstraction of both reactants \citep{Moses2011-Disequilibrium}, thus depending heavily on their abundance in the atmospheres.

\section{1D-RCPE retrievals of WASP-107b} \label{sec: results grid fit}

In this section, we run retrievals based on the grid of 1D-RCPE models.
We start by briefly discussing the main molecular features in cloud-free transit spectra (Sect. \ref{subsec: gridfit-TransmissionSpectra}).
In Sect. \ref{subsec: gridfit-GrayClouds}, we fit with a simplified `grey` cloud set-up for transit observations below 5 $\um$ to gradually build up our understanding of the 1D-RCPE retrieval method.
Afterwards, in Sect. \ref{subsec: gridfit-ComplexClouds}, we combine all information of the previous sections and run retrievals with complex non-grey clouds on all available observations.
For both Sect. \ref{subsec: gridfit-GrayClouds} and Sect. \ref{subsec: gridfit-ComplexClouds}, we present a nominal case in which the results serve as a benchmark.

\subsection{Transmission spectra}
\label{subsec: gridfit-TransmissionSpectra}
\begin{figure*}[h]
	\centering
	\includegraphics[width=17cm]{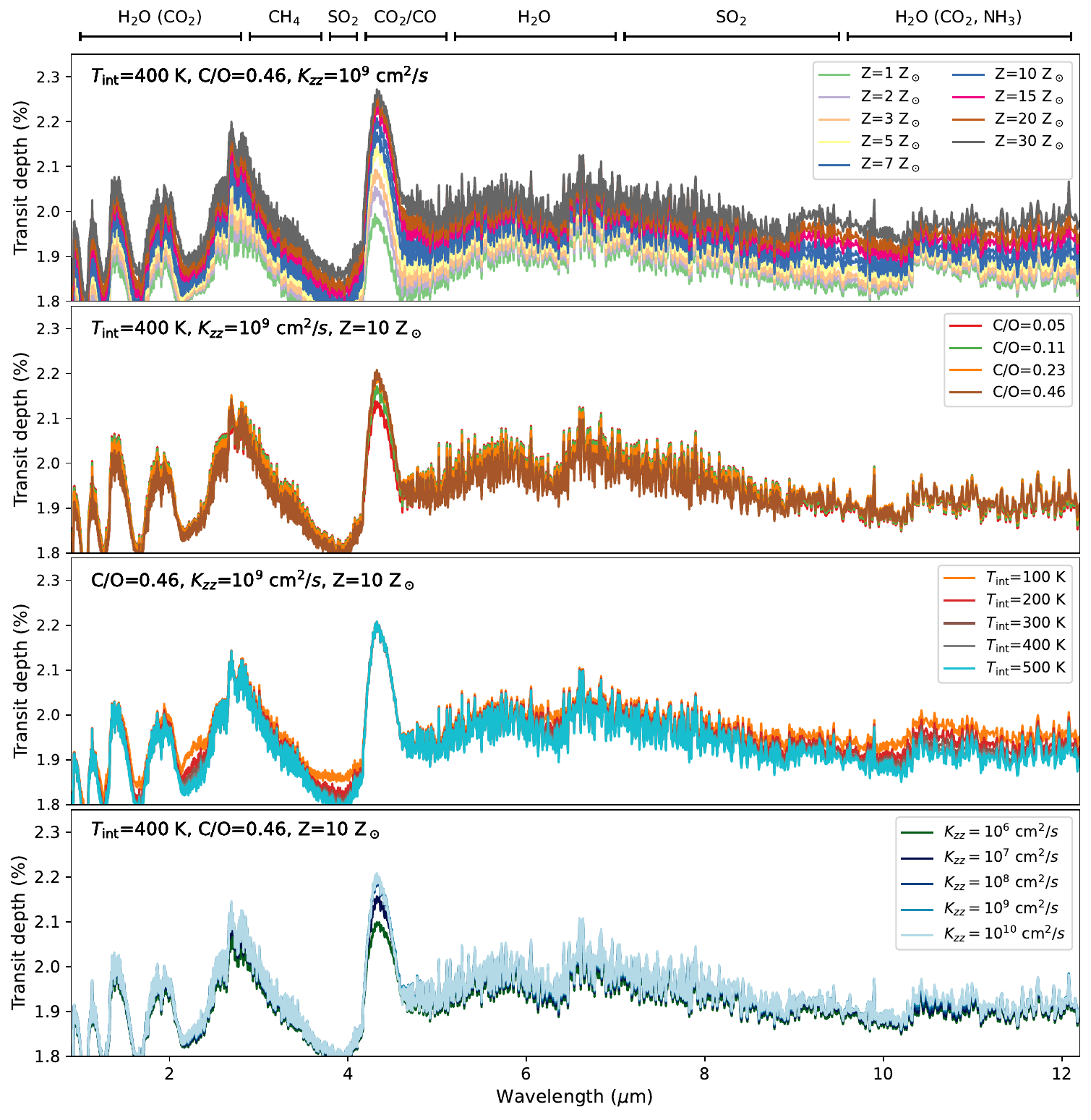}
	\caption{Cloud-free synthetic spectra of our grid of 1D-RCPE models for WASP-107b computed with petitRADTRANS. The top row indicates the dominant opacity sources, with secondary contributors in brackets. } 
	\label{fig: Spectra_grid_plot}
\end{figure*}

With a thorough understanding of the chemical processes and pathways, we are equipped to transform these models into observables.
As is described in Sect. \ref{sec: methods}, we computed transmission spectra of our models on the fly during the fitting routine.
Fig. \ref{fig: Spectra_grid_plot} shows cloud-free transmission spectra that probe the parameter space of our grid.
\newline

\noindent
A number of molecular features show a clear variation in amplitude.
The spectra are dominated by several broad \ce{H2O} bands, a distinct \ce{SO2} bump around $4 \um$ and a double \ce{SO2} feature at 7-9 $\um$, a \ce{CO2} feature at $4.3 \um$ (and contribution at $2.8 \um$), a small \ce{CO} bump around $4.6 \um$, and a narrow-peaked \ce{CH4} feature at $3.4 \um$.
Note that the latter molecule has broad bands and can dominate the spectra in more methane-rich models.
%
The VMRs of \ce{CO2}, \ce{CO}, \ce{H2O}, and \ce{SO2} all scale with metallicity to a certain degree and can therefore provide more opacity in the observable layers at higher metallicity.
At the same time, a high-metallicity atmosphere is more compact due to the increased mean molecular weight, which tends to reduce feature amplitudes as the scale height decreases.
Although both effects counteract each other, the upper panel of Fig. \ref{fig: Spectra_grid_plot} shows a clear positive correlation between feature amplitude and metallicity.
The opposite is true for \ce{CH4} (3.4 $\um$), for which the VMR decreases with metallicity (also see Fig. \ref{fig: Chem_grid_overview_CH4_SO2_NH3}) but the feature strength remains the same for all considered metallicities.
\\

%
%
\noindent At low C/O, less \ce{CH4} is present in the atmosphere and more oxygen is stored in \ce{SO2} and \ce{H2O}, compared to \ce{CO} and \ce{CO2}.
As a result, a sub-solar C/O gives a weaker \ce{CH4} feature and more pronounced \ce{SO2} and \ce{H2O} bands.
Models with high $\Tint$ generally have less \ce{CH4} and \ce{NH3}, which is directly visible in Fig. \ref{fig: Spectra_grid_plot}.
The \ce{CH4} feature at 3.4 $\um$ diminishes in amplitude at higher $\Tint$.
In general, we do not see substantial contribution of \ce{NH3} in these cloud-free spectra, with the exception around $10.5 \um$ in the models with the lowest $\Tint$.
\newline

%
%
\noindent Finally, we briefly discuss the bottom panel of Fig. \ref{fig: Spectra_grid_plot}, where different values of $\Kzz$ are explored.
While higher $\Kzz$ results in deeper quenching, and an overall lower amount of \ce{CH4} in the bulk atmosphere, the effect on the spectral feature at $3.4 \um$ is small.
The \ce{SO2} features around $7-9 \um$ scale with $\Kzz$, which is easily explained by its VMR in the upper layers.
At higher $\Kzz$, vertical mixing is more efficient in replenishing the upper layers with \ce{SO2} as it competes with the photochemical destruction.
Given the high transit depth of these \ce{SO2} features, it is sensitive to exactly these upper layers in the chemistry models, which explains its behaviour in the spectrum.
As we shall see later, the same mechanism can affect the $3.4 \um$ feature of \ce{CH4}, especially in the presence of a high altitude cloud deck that blocks the bulk \ce{CH4} content below from forming a stronger spectral feature.
We emphasize that Fig. \ref{fig: Spectra_grid_plot} was computed for cloud-less spectra, whereas the next sections include clouds that substantially affect the (amplitude of) visible molecular features.

\subsection{Retrievals with grey clouds on transit spectra below 5 $\um$} \label{subsec: gridfit-GrayClouds}
We proceed by fitting our grid of 1D-RCPE models to HST \citep{Kreidberg2018waterw107b} and JWST's NIRCam \citep{Welbanks2024-High-InternalHeatFlux} observations of WASP-107b. 
In this wavelength range, we can highly simplify the cloud treatment and avoid the discussion on more complex (silicate) clouds for now.
The most simple parametrization of a cloud deck can be achieved by applying a grey opacity with a base on certain pressure ($\Pbase$).
We ran our fitting routine (described in Sect. \ref{subsec: Ultranest}) on JWST's NIRCam and HST data and allow for an offset between the two ($\offsetHST$).
We included the full range of grid parameters for Z (1 to 30 $\Zsun$), C/O (0.05 to 0.46), and $\Tint$ (100 to 500 K), with the exception of $\Kzz$ (\num{e7} to \num{e10} $\cmsquareds$).
We excluded models with $\Kzz = \num{e6} \cmsquareds$ as these atmospheres are not always quenched in our grid, which makes a linear interpolation between such models inconsistent.
Finally, we fitted a reference pressure ($\Pref$) at the planetary radius, which gives a total of seven free parameters (Z, C/O, $\Tint$, $\Kzz$, $\Pref$, $\offsetHST$, and $\Pbase$) with flat priors that were fitted for.

\begin{figure*}[h]
	\centering
	\includegraphics[width=17cm]{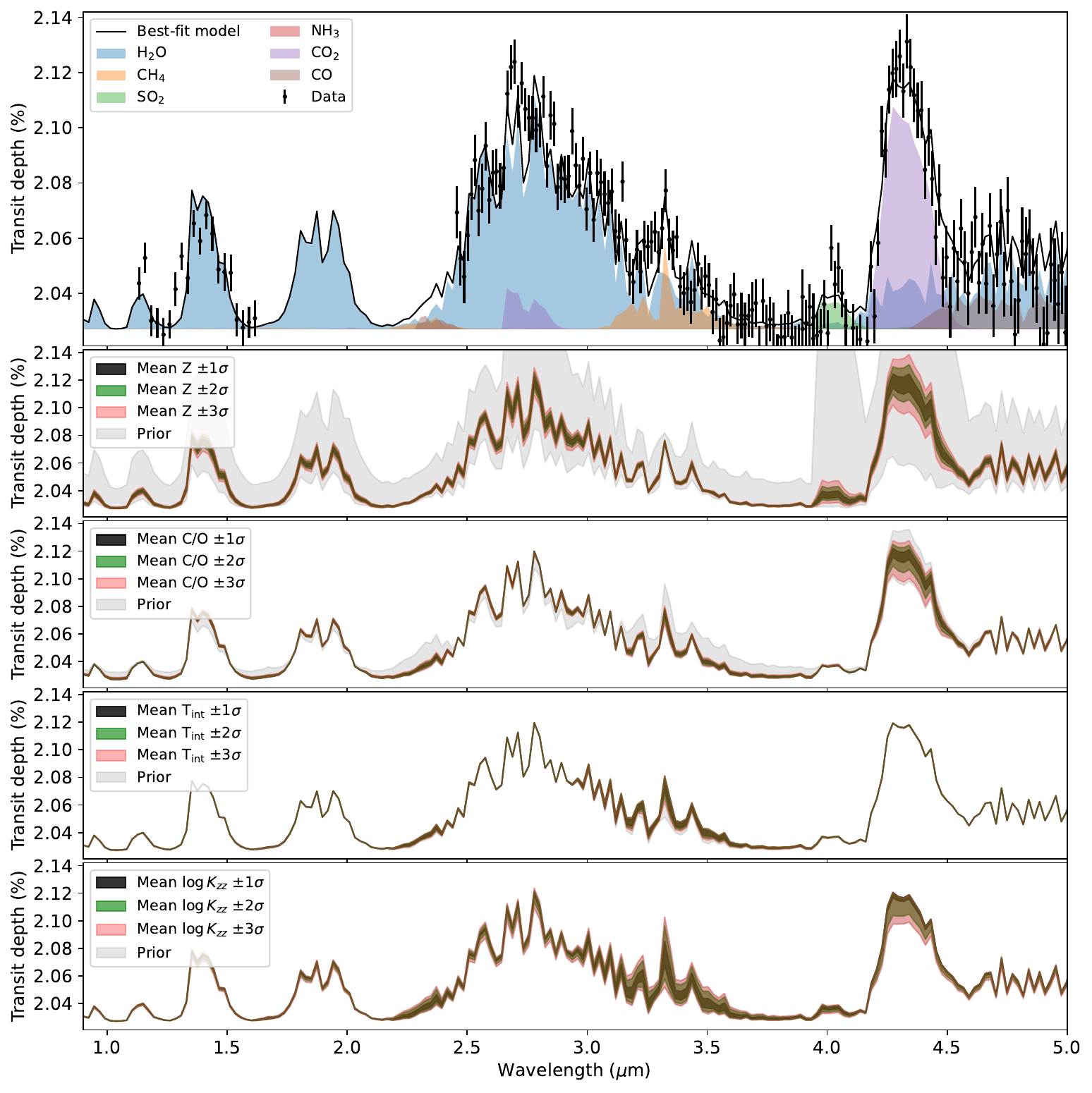}
	\caption{Best-fit spectrum of Case 1 with molecular contributions (upper panel). The bottom panels show spectra within $1 \sigma$, $2 \sigma$, and $3 \sigma$ uncertainty levels of the mean posterior values. The data plotted in the upper panel consists of HST WFC3 \citep{Kreidberg2018waterw107b} and JWST NIRCam \citep{Welbanks2024-High-InternalHeatFlux}.
	}
	\label{fig: Ultranest-HSTNIRCam_FR_GC_bestfit}
\end{figure*}

\subsubsection{Results of the nominal retrieval: Case 1}
Fig. \ref{fig: Ultranest-HSTNIRCam_FR_GC_bestfit} shows the best-fit spectrum of our nominal retrieval (Case 1), summarized in Table \ref{tab: OverviewUltranest}.
By analysing the order of convergence, we see that the metallicity Z = $3.77_{-0.39}^{+0.37} \,\, \Zsun$ and $\log \Pbase$ (bar) = $-3.54_{-0.06}^{+0.07}$ are constrained the easiest, together with offset parameters $\Pref$ and $\offsetHST$.
Upon closer examination (Fig. \ref{fig: Ultranest-HSTNIRCam_FR_GC_cornetplot}), a degeneracy exists between Z, $\Pref$, and $\Pbase$ due to the dominant presence of \ce{H2O}, both due to the amount of spectral features and high VMR, and the strong feature of \ce{CO2} at $4.3 \um$.
Depending on the metallicity, and thus overall \ce{H2O} and \ce{CO2} content, one can vary both $\Pbase$ and $\Pref$ to correctly match the scale height of these features.
Note that an increased metallicity affects the spectrum in two ways.
First, the amplitude of the many \ce{H2O} features (and \ce{CO2}) will scale with their abundance, which can be fitted with a vertical offset (controlled by $\Pref$) to match the top of the feature and $\Pbase$, which blocks the layers below.
Second, an increased metallicity also implies a higher overall mean molecular weight.
This effectively decreases the pressure scale height and hence the physical extent of the atmospheric layers through which star light is absorbed.
This reduces the amplitudes of all spectral features, including those of \ce{H2O} and \ce{CO2}, and has a similar effect to fitting a high-altitude cloud deck (low $\Pbase$). 
It is the $4 \um$ \ce{SO2} feature that breaks this degeneracy, despite slightly underestimating the transit depth in this region, as \ce{SO2} is highly sensitive to metallicity.
In Sect. \ref{subsec: gridfit-ComplexClouds}, we further discuss the ability of \ce{SO2} features to constrain metallicity.
\\

%
%
\noindent The constraint of $\Pbase$ to low pressures is crucial for the retrieval of $\text{C/O} = 0.13_{-0.02}^{+0.02}$, $\Tint$ = $163_{-44}^{+62}$ K, and $\log \Kzz$ = $8.68_{-0.51}^{+0.56}$ (cgs).
The most straightforward explanation of such low C/O would be that an extremely carbon-poor atmosphere removes excessive \ce{CH4} that would otherwise dominate the full wavelength range.
However, also oxygen-rich species are affected by such low carbon-to-oxygen ratios.
Compared to a solar C/O, there is also less \ce{CO2} and more \ce{H2O}, which are both dominant contributors in these spectra.
Fitting the relative proportion between both \ce{H2O} and \ce{CO2} features cannot be done by adjusting the general metallicity.
Instead, one must lower the carbon-to-oxygen ratio to downscale the \ce{CO2} feature with respect to the \ce{H2O} bands, which explains the preference towards such low C/O value.
\\

%
\noindent So far, we have not discussed the 3.4 $\um$ \ce{CH4} feature that is fitted nicely by our retrieval.
Other than a sub-solar C/O, we stated in Sect. \ref{subsec: Carbon-chemistry} that a combination of high $\Tint$ and high $\Kzz$ causes deep quenching and a low \ce{CH4} abundance (Fig. \ref{fig: CH4_Quenching}).
However, we fit our grid to a very low $\Tint$, which leaves a large amount of methane in the atmosphere (see Fig. \ref{fig: Chem_grid_overview_CH4_SO2_NH3}).
Fig. \ref{fig: Ultranest-HSTNIRCam_FR_GC_bestfit} also shows the very limited impact of $\Tint$ on the spectrum for this combination of Z, C/O and $\Pbase$.
We find no substantial difference (< 20 ppm) across the full prior range of considered intrinsic temperatures, a very peculiar result considering our findings in Sect. \ref{subsec: Carbon-chemistry} (Fig. \ref{fig: CH4_Quenching}).
However, by fitting the cloud deck at $\Pbase \simeq 10^{-3.54} $ bar, one effectively shields the layers below where the bulk methane is present.
One still needs to fit the 3.4 $\um$ methane feature in some way, which is achieved via the vertical mixing parameter $\Kzz$.
Although a larger value for $\Kzz$ implies deeper quenching, and thus a lower quenched VMR, it has a different effect in the upper atmosphere.
There, it competes with photochemical destruction to mix additional \ce{CH4} in the upper layers.
Given the high-altitude cloud deck that is retrieved exactly around these layers, the retrieval algorithm prefers models with high $\Kzz$ to provide the necessary amount of methane that can fit the 3.4 $\um$ feature.
This is also in agreement with a lower $\Tint$, as this sets the bulk amount of methane beneath the cloud deck, which is mixed upwards.
Note that this explanation does not contradict our findings of Sect. \ref{subsec: Carbon-chemistry}, as the situation is drastically affected by the cloud deck.
This means that, in this particular case, parameters $\Tint$ and $\Kzz$ were fitted to maximize the \ce{CH4} feature at $3.4 \um$ instead of minimize it.

\subsubsection{Sensitivity analysis: Cases 2, 3, 4, 5, and 6} \label{subsubsec: Gray-sensitivity}
We continue with the same retrieval set-up, but now we systematically fix a parameter beforehand and exclude it from the run.
In this way, we enhance our understanding of the 1D-RCPE retrieval method and demonstrate how our results are influenced by variations in the specific model parameters.
The results of these retrievals are summarized in Table \ref{tab: OverviewUltranest}.
When assuming $\text{Z} = 10 \, \Zsun$ (Case 2) based on previous findings \citep{Welbanks2024-High-InternalHeatFlux, Dyrek2023so2SilicateClouds}, we find a different result compared to our nominal retrieval (Case 1).
Due to the strong presence of the \ce{CO2}, \ce{SO2}, and \ce{H2O} features at this metallicity, the cloud deck was placed at a higher altitude (lower $\Pbase$) to reduce their amplitudes. 
Note that the reference pressure, $\Pref$, was adjusted accordingly, being degenerate with $\Pbase$ and metallicity.
Our retrieval constrains $\Kzz$ to a value close to the lower boundary of the prior (< $10^{7.5}$ $\cmsquareds$).
This is again a peculiar result given the fact that previously we found that the high cloud deck requires a higher $\Kzz$ value to mix sufficient amounts of \ce{CH4} into the upper layers to form a visible spectral feature.
However, the same upward mixing also affects other species such as \ce{H2O}, \ce{CO2}, and in particular \ce{SO2}.
After fixing $\Pbase$ based on the high metallicity, the retrieval minimizes $\Kzz$ to further reduce the scale height of these features.
We note that the carbon-to-oxygen ratio is, again, fitted based on the relative amplitude of the \ce{CO2} and \ce{H2O} bands, which explains the extremely low value.
Finally, we note that $\Tint$ remains unconstrained, as the combination of high clouds and low $\Kzz$ prevents any amount of methane from being visible in the upper layers.
The overall fit is of a much lower quality than the nominal model ($\chi^2_{\rm red} = 2.75$ compared to 2.18 for Case 1).
The reason is twofold as (i) no combination of Z, $\Pref$ and $\Pbase$ can sufficiently reduce the 4 $\um$ \ce{SO2} feature and (ii) the low value of $\Kzz$ prevents \ce{CH4} from forming a feature at $3.4 \um$ above the cloud deck.

\noindent Running the retrieval with a fixed solar C/O (Case 3), gives Z = $2.71_{-0.17}^{+0.22} \, \Zsun$, $\Tint = 490_{-15}^{+7}$ K and $\log \Kzz = 7.18_{-0.13}^{+0.22}$ (cgs).
By fixing the C/O beforehand, and thus the relative amplitude between the \ce{H2O} and \ce{CO2} features, a slightly lower metallicity is preferred to fit these features.
This is a consequence of the non-linear relation between the VMR of \ce{CO2} and Z, while \ce{H2O} does scale linearly with metallicity.
Towards higher metallicities, the increase in \ce{CO2} is greater than \ce{H2O}, requiring a much lower C/O to fit the resulting feature amplitudes.
Note that this results in a poor fit of the \ce{SO2} feature, as there is none at such low metallicity in these spectra.
The slightly lower value for Z results in a deeper cloud deck (higher $\Pbase$), as there is less of a need for these high clouds to reduce the amplitudes of the spectral features. 
These relative deep clouds now expose more layers containing a substantial \ce{CH4} content, which in turn explains the extremely high $\Tint$ and low $\Kzz$.
We repeat that, although a low $\Kzz$ results in more shallow quenching (Fig. \ref{fig: CH4_Quenching}), there is less methane mixed in the upper (visible) layers (see Fig. \ref{fig: Chem_grid_overview_CH4_SO2_NH3}).
\\

\noindent Another exercise is to fix $\Tint = 500 $ K (Case 4), a parameter that is typically constrained with great difficulty as it only affects the 3.4 $\um$ \ce{CH4} feature.
The retrieval first finds a solution between the metallicity, $\Pref$, and $\Pbase$, after which it converges to a sub-solar C/O and low $\Kzz$. 
Interestingly, both Z and C/O are much less tightly constrained than before, encompassing the values of Case 1 within the 1 to 2$\sigma$ level.
As a high $\Tint$ does not affect the VMR of \ce{H2O}, \ce{CO2}, or \ce{SO2}, the only difference stems from starting with a fixed $\Tint$ that affects the VMR of \ce{CH4}.
When we fix $\log \Kzz = 10 $ (cgs) in Case 5, we find results that closely resemble our nominal model with no fixed parameters (Case 1).
Do note the exception that a slightly higher intrinsic temperature is fitted for, which removes additional \ce{CH4}, as more material gets mixed in the upper observable layers as a result of the higher $\Kzz$.
The BIC values for Case 4, and especially Case 5, are close to that of the nominal model (Case 1).
This confirms that $\Tint$ and $\Kzz$ are less important for the retrieval outcome compared to Z and C/O, as they are more difficult to constrain.
\\~\\
Finally, we fixed the cloud deck to $\Pbase = \num{e-3}$ bar (Case 6), which is deeper in the atmosphere than has been previously fitted for.
As usual, this directly affects the metallicity (and $\Pref$), which was now fitted to a relatively low value of Z $= 1.74_{-0.08}^{+0.08} \, \Zsun$ as there was no high cloud deck to reduce the scale height of the spectral features.
We find a slightly higher C/O compared to other runs, which is a result of fitting a lower Z that affects the relative VMR of \ce{CO2} with respect to \ce{H2O}.
Again, this lower Z affects the quality of fit as the \ce{SO2} bump at 4 $\um$ is not fitted well.
The deep cloud deck fails to block the \ce{CH4}-rich layers.
Therefore, a combination of high $\Tint$ and low $\Kzz$ minimizes the \ce{CH4} opacity contribution to the spectrum, explaining the final results.

\subsubsection{A wavelength-dependent opacity slope: Case 7} \label{subsubsec: Gray-Scattering}
A consistent result among all runs is the very low C/O value.
The retrievals achieve these values by trying to reduce the \ce{CO2} feature ($4.3 \um$) in amplitude with respect to the \ce{H2O} band ($2.5 - 3.0 \um$) or vice versa.
Rayleigh scattering can introduce a slope towards the optical part of the transit spectrum, effectively altering the fit for the HST data (and thus $\offsetHST$).
Given the detection of larger silicate droplets in WASP-107b \citep{Dyrek2023so2SilicateClouds}, a non-Rayleigh type of short-wavelength absorption could be present that follows a much more shallow wavelength-dependence.
In Case 7, we fit for a free slope and find $\gammascatt = -0.75_{-0.09}^{0.09}$, which is more indicative of Mie scattering caused by mid-sized to larger aerosols.
The effect on C/O is minimal, however, as we retrieve a similar value as before.
We do obtain a better fit compared to the nominal model (Case 1), as well as a better BIC, indicating that this retrieval manages to explain the data better.
Note that $\Pbase$ loses its meaning as the opacity source with the shallow slope now replaces the grey cloud deck ($\Pbase$).
Therefore, in future simulations with a fitted $\gammascatt$, we do not fit for $\Pbase$.

\subsubsection{Information content of HST and JWST NIRCam data: Cases 8 and 9} \label{subsubsec: Gray-Informationcontent}

Finally, we briefly consider different combinations of datasets in these retrievals with grey clouds, excluding the JWST MIRI data of WASP-107b for now.
A fit to the HST dataset alone (Case 8) only constrains a high cloud deck, as no features other than of \ce{H2O} are visible.
In particular, the metallicity remains unconstrained, which provides further evidence that features from only \ce{H2O} are not sufficient to constrain Z.
Furthermore, fitting to the NIRCam dataset separately (Case 9) gives similar results as to fitting it in combination with HST data (Case 1).
We conclude that, as expected \citep[e.g.][]{Edwards2023Exploring}, the retrieval needs a feature-rich spectrum such as JWST NIRCam to constrain atmospheric properties.

\subsection{Retrievals with non-grey clouds on transit spectra up to 12.2 $\um$}
\label{subsec: gridfit-ComplexClouds}

\begin{figure*}[h]
	\centering
	\includegraphics[width=17cm]{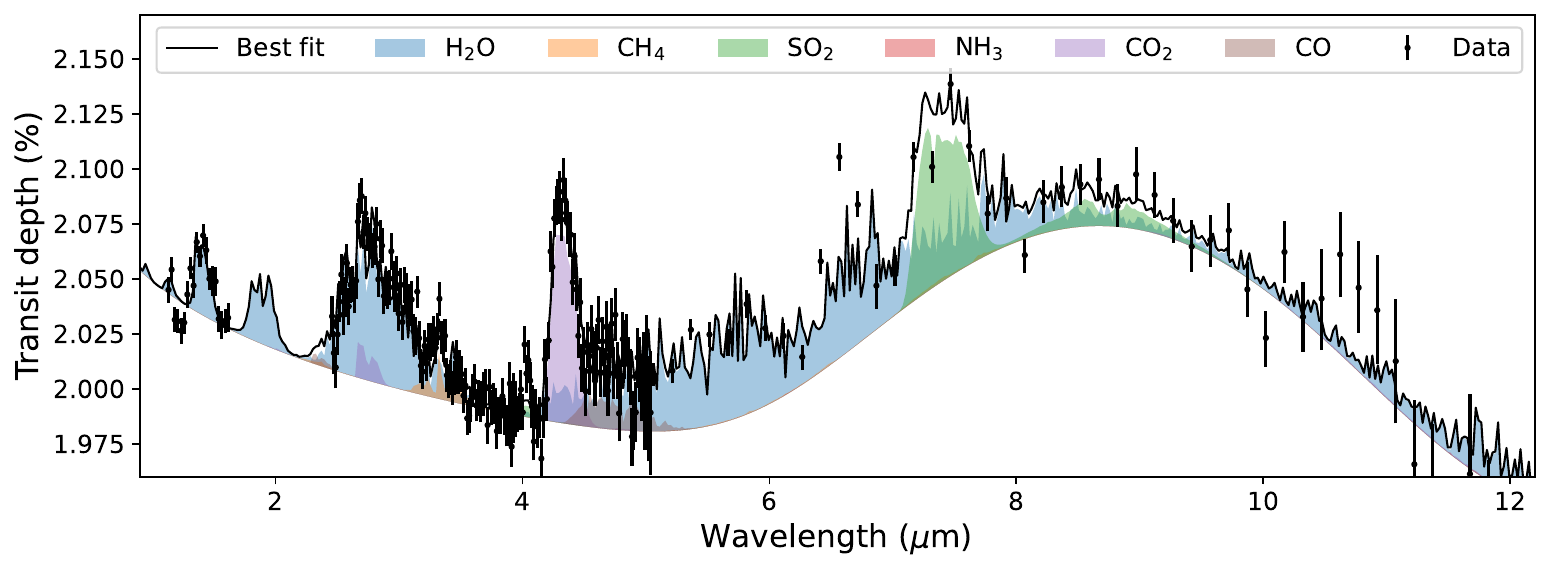}
	\caption{Best-fit spectrum for Case 10. The data plotted in the bottom panel consists of HST WFC3 \citep{Kreidberg2018waterw107b}, JWST NIRCam \citep{Welbanks2024-High-InternalHeatFlux}, and JWST MIRI/LRS \citep{Dyrek2023so2SilicateClouds}.}
	\label{fig: Bestfit_LW}
\end{figure*}

Observations with the JWST MIRI instrument (LRS) provide us with additional wavelength coverage up to ${\sim} 12.2 \um$.
Given the additional opacity source between 8 and 10 $\um$, stemming from (silicate) clouds \citep{Dyrek2023so2SilicateClouds, Welbanks2024-High-InternalHeatFlux}, we ran retrievals with non-grey cloud parametrizations described in Sect. \ref{subsec: cloudparams}.
We included uniform priors that span the entire range of the four grid parameters (Z, C/O, $\Tint$, $\Kzz$) with the exception of $\log \Kzz = 6 $ (cgs) as we only want to include quenched atmospheres.
Next to an offset with HST data ($\offsetHST$), we also fitted an offset with the NIRCam data ($\offsetNIR$) in reference to the MIRI data.
Together with the model offset ($\Pref$), this makes a total of seven free parameters without the cloud treatment.

\subsubsection{Results with Gaussian clouds: Cases 10, 11, 12, 13, and 14}

\noindent We started by adopting the cloud parametrization of \citet{Welbanks2024-High-InternalHeatFlux}, which introduce three more fitting parameters ($\kappaWelbanks$, $\lambda_0$, $\omega$) as we fixed $\xi = 0$ after initial testing to reduce the computational cost.
Additionally, we included a wavelength-dependent opacity slope as described in Eq. \ref{eq: Scattering parametrization}, which introduces two additional fitting parameters to the retrieval ($\kappascatt$, $\gammascatt$).
Note that we do not fit for a grey cloud deck ($\Pbase$) if we consider $\gammascatt$ in the retrieval.
\\

\noindent Figs. \ref{fig: Bestfit_LW} and \ref{fig: CornerPlot_LW} show the best fit spectrum marginal posterior distributions of this retrieval (Case 10).
Overall, we achieve a good fit for all datasets with the exception of the noisy region above 10 $\um$.
At $4 \um$, we find that our best-fit model contains mostly \ce{SO2} opacity, although it does not reproduce the narrow peak in the data completely.
As is established in Sect. \ref{subsec: gridfit-GrayClouds}, the metallicity is degenerate with the model offset ($\Pref$) and cloud opacity (e.g. $\Pbase$) when fitting to the \ce{H2O} and \ce{CO2} features in the transit spectrum below $5 \um$.
Only the $4 \um$ feature of \ce{SO2} helps constrain the metallicity as higher values for Z (and thus higher VMR of \ce{SO2}) would result in a \ce{SO2} feature with too high amplitude.
Interestingly, the inclusion of two broad \ce{SO2} features around $7-9 \um$ does not change the retrieved metallicity ($\text{Z} = 3.78_{-0.53}^{+0.53} \,\, \Zsun$) significantly compared to previous retrievals with grey clouds ($\text{Z} = 3.77_{-0.39}^{+0.37} \,\, \Zsun$, see Table \ref{tab: OverviewUltranest}).
This could indicate that any of the \ce{SO2} features are a robust indicator of atmospheric metallicity, or that the Gaussian cloud feature is degenerate with the \ce{SO2} features at $7-9 \um$, and thus metallicity is constrained only on the $4 \um$ \ce{SO2} feature.
\\

\noindent If \ce{SO2} is a strong indicator of the metallicity in a photochemically active atmosphere, its features in the MIRI data should cause the retrieval to converge to a similar value for Z.
Indeed, re-running the retrieval without HST and NIRCAM data (Case 14, with $\Pbase$ instead of $\gammascatt$) results in Z  = $3.16_{-0.43}^{+0.78} \,\, \Zsun$, which is within the $ 1\sigma$ level as before (Case 10). 
Further investigation confirms that the quality of fit decreases with a fixed higher metallicity (see BIC of Case 11 with Z = 10 $\Zsun$ in Table \ref{tab: OverviewUltranest_Gaussianclouds}).
We conclude that the \ce{SO2} features, either at 4 $\um$ or $7-9 \um$, are robust features to constrain the metallicity.
In Sect. \ref{sec: discussion}, we further discuss their dependence on model assumptions such as the temperature in the upper layers.
\\

\noindent The inclusion of MIRI data introduces features of two oxygen-bearing species (\ce{H2O} and \ce{SO2}) and (a lack of) one carbon-bearing molecule (\ce{CH4}), which can further impact the C/O.
We find that the nominal retrieval (Case 10) has difficulties constraining C/O (next to $\Tint$ and $\Kzz$) compared to other parameters such as Z, $\Pref$, and the cloud properties.
Similar to our findings before (Sect. \ref{subsec: gridfit-GrayClouds}), we find that the easy constraints on cloud (and scattering) opacity affect the further convergence of C/O.
As the high-altitude layer of aerosols shields the layers with substantial \ce{CH4}, and thus the formation of any features in the spectrum, the C/O ($0.20_{-0.02}^{+0.03}$) is constrained by the relative amplitude of the $4.3 \um$ \ce{CO2} feature with respect to features of oxygen species (e.g. \ce{H2O} and \ce{SO2} in NIRCAM).
Within the MIRI data, lowering the C/O effectively reduces the amplitude of both \ce{SO2} and \ce{H2O} feature as no carbon-species has a strong feature.
Therefore, we find that the C/O is still mainly controlled by the 4.3 $\um$ \ce{CO2} feature.
To further assess the sensitivity of C/O to the scattering treatment, we run the same retrieval without scattering and find a marginally lower C/O of $0.15_{-0.02}^{+0.02}$ (Case 13, Table \ref{tab: OverviewUltranest_Gaussianclouds}).
Furthermore, a simulation with a fixed solar C/O gives a satisfactory fit (see BIC of Case 12 in Table \ref{tab: OverviewUltranest_Gaussianclouds}), indicating that an adjusted low-wavelength opacity slope can reduce the need to fit a low C/O.
We carefully conclude that the above indicates a degeneracy between C/O and the wavelength-dependent opacity slope. 
\\

\noindent Parameters $\Tint$ and $\Kzz$ are least constrained by the retrieval.
We find a relatively high intrinsic temperature of $369_{-70}^{+60} \, \text{K}$ and intermediate $\log \Kzz$ of $8.25_{-0.68}^{+0.64}$ (cgs).
As was previously established, a high-altitude cloud deck shields the \ce{CH4}-rich layers and prevents it from forming features in the observable spectrum.
Once again, this hypothesis is confirmed by finding no substantial difference (< 15 ppm) between model spectra across the full range of $\Tint$, while otherwise adopting the best-fit values.
The \ce{CH4} feature at $3.4 \um$ is fitted by increasing $\Kzz$ so that additional \ce{CH4} is mixed upwards and reaches above the cloud deck.
While $\Tint$ mostly affects \ce{CH4} features in the transmission spectrum, we do note that $\Kzz$ also affects the abundances of \ce{SO2} and \ce{CO2} (e.g. Fig. \ref{fig: Chem_grid_overview_CH4_SO2_NH3} and Fig. \ref{appendixfig: Chemistry_grid_plot_HCN_CO2_C2H2_SO_H2S}), and therefore also the scale height of their spectral features.
\\

\noindent Finally, we note that none of the above fits contain a significant contribution from \ce{NH3}, a species detected by \citet{Welbanks2024-High-InternalHeatFlux} at $3 \um$ (and $10.5 \um$ to a lesser extent) in WASP-107b.
Although we do not obtain a good fit overall above $10 \, \um$ with our models, the region around $3 \, \um$ is well fitted by mostly \ce{H2O} opacity instead of \ce{NH3}. 
As is discussed in Sect. \ref{subsec: Nitrogen-chemistry} and shown in Fig. \ref{fig: Chem_grid_overview_CH4_SO2_NH3}, \ce{NH3} is most abundant in atmospheric models with low $\Tint$.
Similar to \ce{CH4}, high altitude clouds could obscure \ce{NH3}-rich layers and limit the amplitude of any potential spectral features.
Meanwhile, at a high $\Kzz$ additional material could be mixed upwards above the cloud deck and become visible in the transit spectrum.
However, we do not find substantial opacity contributions of \ce{NH3} in cloud-free models (Fig. \ref{fig: Spectra_grid_plot}).
To further investigate the impact of \ce{NH3}, we run a retrieval with Gaussian clouds and a wavelength-dependent opacity slope (Similar to Case 10) without \ce{NH3} opacity in the radiative transfer computation.
We find that this does not affect the retrieval outcome and conclude that \ce{NH3} opacity does not play a role in any of the above retrievals.

\subsubsection{Retrievals with silicate condensate clouds: Cases 15, 16, and 17}
Given the detection of silicate condensates in the MIRI data \citep{Dyrek2023so2SilicateClouds}, we run retrievals that use optical properties of \ce{SiO}, \ce{SiO2}, or \ce{MgSiO3} to explore their effect on other fitting parameters.
As is detailed in Sect. \ref{subsec: cloudparams}, this introduces four more parameters ($\Xbase$, $\Pbase$, $\fsed$, $\Kzzcloud$) as we fix $\xsigma = -1$.
Next to the four grid dimensions (Z, C/O, $\Tint$, $\Kzz$) and three offsets ($\Pref$, $\offsetHST$, $\offsetNIR$), this totals 11 fitting parameters.
Note that the implementation of particle size and mass fraction distribution removes the need for separate parametrized scattering.
As expected, we retrieve metallicities and carbon-to-oxygen ratios that are in line with previous runs, confirming that both these parameters are not sensitive to the way we choose to parametrization clouds in our set-up.
In contrast to our results with Gaussian clouds, we now find lower intrinsic temperature (${\lesssim} \, 250$ K).
In line with our previous findings, a low $\Tint$ is caused by the altitude at which the aerosols become optically thick and whether or not this exposes the \ce{CH4}-rich layers.
Another notable difference with the Gaussian cloud parametrized retrievals is the quality of fit, which is much lower in this case.
Closer inspection reveals that the JWST MIRI data is not properly fitted for (Fig. \ref{appendixfig: Silicate_clouds}), which confirms that a single silicate species cannot explain the broad $8$ to $10 \um$ cloud feature \citep{Dyrek2023so2SilicateClouds}.
Although determining which (combination of) silicate cloud species or other condensates 
can fit the data better is out of the scope of this study, we briefly discuss the debated presence of silicate clouds in Sect. \ref{subsec: SilicateClouds?}.

\section{Discussion} \label{sec: discussion}
In this section we discuss the implications of our results for WASP-107b.
First, we discuss our findings on the elemental composition (Z and C/O) in Sect. \ref{subsec: Retrieved composition} before discussing the reason why our retrievals cannot constrain $\Tint$ and $\Kzz$ properly in Sect. \ref{subsec: High-altitude aerosols}.
Finally, we review the presence of high-altitude clouds that consist of silicates particles in Sect. \ref{subsec: SilicateClouds?}.

\subsection{Retrieved composition of WASP-107b} \label{subsec: Retrieved composition}

We consistently find metallicities of $3-5 \, \Zsun$ in retrievals that achieve a good fit.
As is discussed extensively in Sect. \ref{sec: results grid fit}, the metallicity is constrained based on the \ce{SO2} features at $4$ and $7-9 \um$ as \ce{SO2} production is dependent on the available sulphur (\ce{H2S}) and oxygen reservoirs (\ce{H2O}) (see Sect. \ref{subsec: Sulphur-chemistry}).
Sulphur dioxide effectively breaks the degeneracy between metallicity (based on features of \ce{H2O}, and \ce{CO2}), model offset (fitted $\Pref$ at $\rplanet$), and top of the cloud deck ($\Pbase$).
Our derived metallicities are lower than values reported in the literature, which range between 10 to 18 $\Zsun$ \citep{Welbanks2024-High-InternalHeatFlux} and $43 \pm 8 \, \Zsun$ \citep{Sing2024-WarmNeptunes}.
As \ce{SO2} is a photochemical product, its VMR profile is determined by (i) the SED emitted by the host star (in particular the UV), 
and (ii) the temperature structure \citep{Tsai2023PhotochemicallyproducedSO2, Dyrek2023so2SilicateClouds, deGruyter2024}.
We note that \citet{Welbanks2024-High-InternalHeatFlux} use the same SED (HD 85512 from the MUSCLES survey) as we do in this study.
Therefore, we attribute part of the difference in derived metallicity to their lower upper-layer temperature (450 - 550 K) with respect to ours (650 - 750 K).
At lower temperature, the thermochemical conversion from \ce{H2O} into \ce{OH} (Re. \ref{Reaction: Tsai-H2O+H->OH+H2}) becomes slower, which suppresses the efficiency of sulphur oxidization (e.g. Re. \ref{Reaction: Tsai-SO+OH->SO2+H}).
Hence, a higher metallicity is required to produce sufficient \ce{SO2} in their models.
\begin{figure}[h]
	\resizebox{\hsize}{!}{\includegraphics{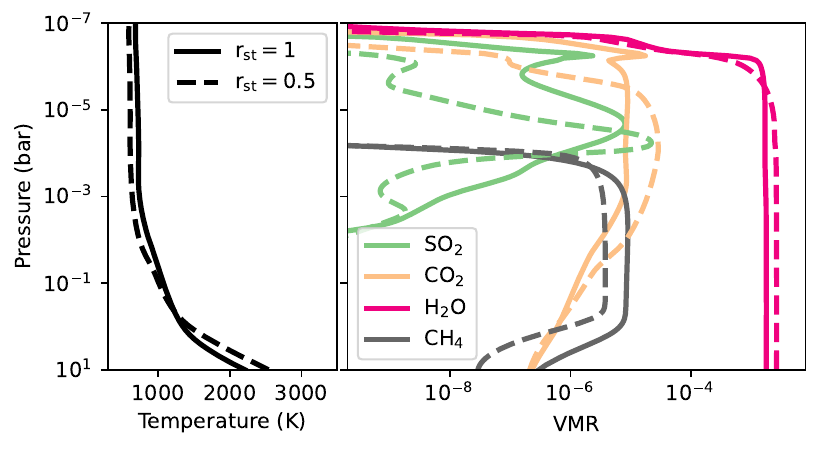}}
	\resizebox{\hsize}{!}{\includegraphics{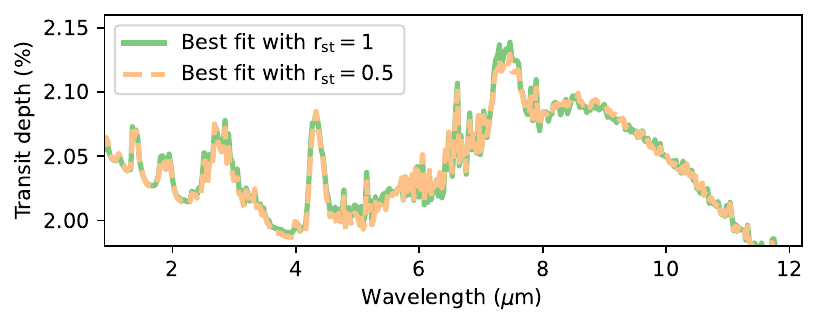}}
	\caption{
		Best-fit PT profile (upper left), VMRs of key species (upper right), and model spectra (bottom) of retrievals Case 19 and Case 10 that have Gaussian clouds and a free wavelength-dependent opacity slope.
		The grid of models used for Case 19 are approximately 100 K cooler in the isothermal layers (r$_{\rm st}  = 0.5$) compared to those used for Case 10 (r$_{\rm st}  = 1$).
		Note that both retrievals converge to a different set of grid parameters (Z, C/O, $\Tint$, $\Kzz$), which affects the VMRs and PT profiles, but a fit of similar quality can be achieved.
		}
	\label{fig: Grid6Comparison}
\end{figure}

\noindent The strong sensitivity of \ce{SO2} to temperature inspires us to revisit an assumption made in the construction of PT profiles (Sect. \ref{subsec: picaso}).
By fixing r$_{\rm st}  = 1$ (effectively a dayside average PT structure), we computed a profile with a quasi-isothermal upper atmosphere of $650 - 750$ K while setting r$_{\rm st}  = 0.5 $ (planet-wide average) would lower this by ${\sim}100$ K.
To test the effect of this assumption on our results, we ran an additional grid of 900 models with these slightly cooler PT profiles.
We then ran retrievals using the same set-ups as Case 1 (Table \ref{tab: OverviewUltranest}) and Case 10 (Table \ref{tab: OverviewUltranest_Gaussianclouds}); we present the results in Table \ref{tab: grid6Comparison}.
We find metallicities of $3.92_{-0.36}^{+0.34} \, \Zsun$ (Case 18) and $4.49_{-0.27}^{+0.32} \, \Zsun$ (Case 19), respectively, which indicates only a marginal increase in retrieved metallicity compared to our initial retrievals with r$_{\rm st}  = 1$.
Fig. \ref{fig: Grid6Comparison} shows the best-fit spectrum of Case 19 and Case 10, as well as the PT profiles and VMRs of key species.
Despite a general decrease in \ce{SO2} in the upper regions, a similar fit can be achieved by leveraging the $\rplanet - \Pref$ degeneracy and cloud opacity.
This indicates that the predictive power of \ce{SO2} for metallicity indeed depends on the assumed temperature in the upper layers.
What further complicates the correct estimation of the upper layer temperature is the fact that observations do not probe a 1D atmosphere and thus we cannot assume a uniform temperature on the terminator region.
A slight asymmetry between the evening and morning limb has been detected on WASP-107b, resulting from a ${\sim} 100$ K difference \citep{Murphy2024EvidenceforAsymmetry}.
Limb asymmetry not only affects the temperature and chemistry but changes the scale heights as well, which can bias the observed transit spectrum \citep{Caldas2019-Effectsofafully3D, Falco2021-Towardamultidimensional-I, Pluriel2021-Towardamultidimensional-II, Lee2021-3DradiativeTransferfor, Nixon2022-AURA3D}.

\noindent Another consistent finding of our retrievals is the carbon depletion with respect to oxygen (C/O), typically in the range of 0.10 to 0.20.
We find that this parameter is mainly determined to the relative amplitude of the $4.3 \um$ \ce{CO2} feature with respect to broad features of oxygen species, in particular \ce{H2O}.
If the retrieval of this parameter is reliant on a single \ce{CO2} feature, one must question the relatability and investigate the underlying chemistry that produces \ce{CO2}.
As is detailed in Sect. \ref{subsec: Carbon-chemistry}, we find that the VMR of \ce{CO2} is mainly set by thermochemistry and is less sensitive to disequilibrium effects in our models of WASP-107b.
Retrievals with cooler upper layers (r$_{\rm st}  = 0.5$) prefer models with a slightly lower C/O (Table \ref{tab: grid6Comparison}), indicating an even higher carbon depletion is needed to fit the spectra properly.

\subsection{High-altitude aerosols can shield \ce{CH4}-rich layers} \label{subsec: High-altitude aerosols}
We retrieve intrinsic temperatures that span the entire range of our priors (100 to 500 K).
Together with $\Kzz$, our retrievals struggle to constrain $\Tint$ and both parameters often have large uncertainties.
We attribute the large spread in $\Tint$ values to high-altitude clouds that block the layers below from forming spectral features, like the $3.4 \um$ \ce{CH4} feature.
Additionally, we find that strong mixing can mix material above the cloud deck so that a higher $\Kzz$ implies a stronger feature.
This is counter-intuitive from a chemical standpoint, which shows that stronger mixing results in deeper quenching of \ce{CH4} and thus lower VMR (Fig. \ref{fig: CH4_Quenching}).
We conclude that the feature at $3.4 \um$ is fitted by a delicate balance of $\Pbase$ (or equivalent aerosol opacity) and $\Kzz$, while $\Tint$ is less important than expected.
\\

\noindent Previous studies have suggested that a high intrinsic temperature, in combination with deep quenching by efficient vertical mixing, is responsible for the lack of abundant methane in WASP-107b \citep{Kreidberg2018waterw107b, Fortney2020BeyondEq, Dyrek2023so2SilicateClouds, Welbanks2024-High-InternalHeatFlux, Sing2024-WarmNeptunes}.
In particular, \citet{Welbanks2024-High-InternalHeatFlux} constrains $\Tint > 345$ K and $\log \Kzz = 8.4-9.0 \cmsquareds$, which our retrievals reproduce when a similar set-up is used (Case 10, Table \ref{tab: OverviewUltranest_Gaussianclouds}).
Although this paper shows that these results can vary with different modelling choices and retrieval set-ups, we can neither confirm nor reject the high internal temperature scenario for WASP-107b.
A hot interior is consistent with the highly inflated radius and low density, likely caused by tidal inflation \citep{Leconte2010IsTidalHeatingSufficient, Fortney2020BeyondEq, Welbanks2024-High-InternalHeatFlux}, and the suggested presence of high-altitude silicates \citep{Dyrek2023so2SilicateClouds}.
Instead, we argue that hybrid retrieval approaches must be interpreted with caution and their robustness verified against various assumptions.
\\~\\
The delicate balance between $\Tint$, $\Kzz$, and clouds relates directly to the underlying grid of forward models.
If the cloud deck is fitted slightly deeper in the atmosphere, and more \ce{CH4} is exposed to the observable layers, the retrieval tends to maximize $\Tint$ and minimize $\Kzz$, while the opposite is true for a slightly higher cloud deck.
This behaviour could be a consequence of the disconnect between our treatment of disequilibrium chemistry and post-processing of clouds in the radiative transfer.
In particular, it can be expected that the methane homopause is shifted upwards as high clouds act as an additional UV absorber, shielding lower layers from photochemistry and increasing local heating rates \citep{Molaverdikhani2020TheRoleofClouds, SanchezLavega2023Dynamics}.
Additionally, clouds affect the gas-phase chemistry \citep{Mahapatra2017CloudFormation, Helling2019-ExoplanetClouds, Kiefer2024Fully}, which in turn also affects the PT structure \citep{Marley2013-CloudsandHazes, Lee2024DynamicallyCoupled}.
This calls for a general framework that allows feedback between (kinetic) cloud formation, gas-phase disequilibrium chemistry, and radiative transfer.
\\

%
%
%
\noindent Finally, we note that other processes can deplete \ce{CH4} as well.
Zonal quenching can be induced by equatorial mixing \citep{Bell2024NightsideClouds}, although this effect could be small for WASP-107b \citep{Baeyens2021-GridofPseudo2D}.
Furthermore, vertical mixing is treated as eddy diffusion, which assumes length scales that are much smaller than the pressure scale height, thereby mimicking diffusion \citep{Tsaj2017-VULCANaopensource}.
However, given the potentially large interior temperatures and resulting steep temperature-pressure gradients, a large part of WASP-107b's atmosphere could be dominated by convective energy transport.
Therefore, additional large-scale mixing might take place that is not captured by the constant eddy diffusion coefficients \citep{Baeyens2021-GridofPseudo2D}.

\subsection{Silicate clouds at high altitudes} \label{subsec: SilicateClouds?}

\citet{Dyrek2023so2SilicateClouds} report a statistical preference for silicate species \ce{SiO}, \ce{SiO2}, and \ce{MgSiO3} with respect to a simpler cloud set-up.
Particularly notable is the high altitude at which the silicate cloud deck was retrieved, while condensation of such silicates occurs at high temperatures (e.g. 1200 to 1500 K) \citep{Lodders1999IAUS..191..279L, Visscher2010ApJ...716.1060V} and thus much deeper in the atmosphere. 
Given this expected deep condensation, \citet{Welbanks2024-High-InternalHeatFlux} show that vertically extended clouds are required to become visible in the transit spectrum ($\fsed$ < 0.5).
However, the resulting small particles give an unsatisfactory fit to the observations.
\citet{Dyrek2023so2SilicateClouds} find a good fit with large silicate particles ($\fsed$ > 3) so that the question remains how such a compact cloud deck of large aerosols can be suspended at high altitudes, far above the expected condensation region.
One possibility is that efficient vertical mixing can loft particles to high altitudes continuously before they grow large and settle down.
Given the intermediate value for $\fsed$ that \citet{Dyrek2023so2SilicateClouds} find, atmospheres with high $\Kzz$ would require particles to grow large quickly and form a relative compact cloud at such high altitudes.
WASP-107b's low gravity could facilitate this slow settling and maintain high-altitude silicate clouds.
\\

\noindent Although the focus of this study is to assess the robustness of retrieval outcomes against different assumptions, including the cloud parametrization, we briefly put our results in context of the above.
In retrievals with individual silicate cloud species, we find $\fsed$ values roughly between 1.50 and 3, moderate values for $\Kzzcloud$, and relatively high altitudes of the cloud pressure ($\Pbase$).
This supports the case of a high-altitude thin cloud deck composed of moderate-sized particles, although this challenges the hypothesis of strong vertical mixing that replenishes these layers and reduces the sedimentation efficiency.
When fitting with Gaussian clouds and a wavelength-dependent opacity slope \citep{Welbanks2024-High-InternalHeatFlux}, we retrieve $\gammascatt = -0.83_{-0.09}^{+0.09}$ (Case 10).
The absence of a steeper scattering slope (e.g. $\gammascatt = -4$) indicates that sub-micron particles are not dominant in the atmosphere.
Instead, the fitted shallow slope resembles more Mie-type scattering, caused by particles with comparable sizes to the photon wavelength, supporting larger grains of one to several micron.

\section{Summary and conclusion} \label{sec: conclusion}
In this paper, we have conducted an in-depth study on the performance and reliability of 1D-RCPE retrievals on transmission spectra of WASP-107b.
We have built a grid of 900 atmospheric models that take into account radiative heating, convection, vertical mixing, and thermo- and photochemistry and explore a range of different metallicities, carbon-to-oxygen ratios, intrinsic temperatures, and eddy diffusion coefficients. 
We have analysed the behaviour of WASP-107b's gas-phase chemistry against different parameters and mapped important chemical pathways for key species in the atmosphere.
We have subsequently used a nested sampling algorithm to perform Bayesian inference on archival datasets of WASP-107b and assessed the robustness and reliability of our outcomes against different cloud treatments.
\\

\noindent Below, we summarize our main findings on the chemistry in our models of WASP-107b.
The main formation pathway of \ce{SO2} in WASP-107b is through oxidization of atomic sulphur or $\ce{H2S} + 2\text{H}_2\text{O} \rightarrow \text{SO}_2 + 3\text{H}_2$, with variations at a high atmospheric metallicity involving oxidization of di-atomic sulphur (\ce{S2}) and/or direct combination of two \ce{SO} compounds into \ce{SO2}.
Hydroxide radicals are available due to hydrogen abstraction and photolysis of \ce{H2O}. 
Atomic sulphur is typically liberated by hydrogen abstraction of \ce{H2S}.
The production of \ce{SO2} is therefore also sensitive to atmospheric metallicity.
Furthermore, we explain how photolysis of non-hydrogen-bearing species such as \ce{N2}, in the absence of other photon absorbers, can interact with the \ce{H2} reservoir to enable additional \ce{SO2} production.
Regarding \ce{CH4}, we reproduce the well-known \ce{CH4}$-$\ce{CO} conversion pathway via hydroxy-methyl (\ce{CH2OH}) that affects the quenching pressure and VMRs deep in the atmosphere.
We find a negative correlation between the quenched \ce{CH4} content and atmospheric metallicity due to the additional heating of the deep layers.
At high $\Tint$, we find that strong mixing ($\Kzz$) results in deep quenching and decreases the quenched VMR of \ce{CH4}, while the opposite is true for models with low $\Tint$.
Finally, we find that \ce{NH3} quenching occurs much deeper in the atmosphere and is less sensitive to $\Kzz$ due to the steep temperature gradient around the quenching point.
\\

\noindent Below, we summarize our main findings of the 1D-RCPE retrievals.
\begin{itemize}
	\item We find that \ce{SO2} features at 4 and 7-9 $\um$ are most informative to determine the atmospheric metallicity, which is consistently retrieved to be between $3$ and $5 \, \Zsun$ in retrievals that achieve a good fit. This is the case when we consider JWST NIRCam data separately, for JWST MIRI/LRS data if the $10 \um$ silicate cloud feature is taken into account, or a combination of both. We note that the amplitude of the \ce{SO2} features is, next to the stellar SED, dependent on the assumed temperature in the upper layers as this affects the available hydroxide that oxidizes sulphur.
	
	\item In the absence of strong \ce{SO2} features, we find a degeneracy between metallicity, the fitted reference pressure, $\Pref$, and the altitude of the clouds ($\Pbase$). This is because the latter two parameters can control the feature amplitude regardless of the VMRs of dominant absorbers such as \ce{H2O} and \ce{CO2}, which scale with Z.
	
	\item The carbon-to-oxygen ratio (C/O) was fitted to sub-solar values ($\lesssim$ 0.20) based on the relative amplitude of the large $4.3 \um$ \ce{CO2} feature with respect to broad bands of \ce{H2O} (and \ce{SO2}), and much less to the $3.4 \um$ \ce{CH4} feature as the latter is often shielded by high-altitude clouds. This sensitivity highlights the need to properly model the VMR profile of \ce{CO2}, which is sensitive to thermochemistry for WASP-107b and disequilibrium chemistry in some other cases.
	
	\item In general, we find a slightly higher C/O when a wavelength-dependent opacity slope is included that mimics scattering towards the optical transit spectrum. The main reason for this is that such opacity increases the transit depth of the broad \ce{H2O} band around $2.5 - 3.0 \um$ with respect to the 4.3 $\um$ \ce{CO2} feature. Therefore, less carbon depletion is necessary to achieve a good fit to the NIRCam data. We stress that this dependence is weak and only changes our results by less than $1 \sigma$ with respect to a nominal model without such a sloped opacity source.
	
	\item We find that the retrieval has difficulty constraining $\Tint$ and $\Kzz$. Depending on the fitted altitude of the aerosol opacity (e.g. by $\Pbase$, $\kappascatt$, $\gammascatt$), $\Tint$ can even remain unconstrained. The reason for this is that high-altitude aerosols shield the \ce{CH4} layers, preventing models with low $\Tint$ from forming significant methane features in the spectrum other than the one at 3.4 $\um$. 
	In retrievals with deeper clouds, a high intrinsic temperature is retrieved.
	This suggests that exoplanet atmospheres with high-altitude clouds prevent us from constraining the intrinsic temperature.
	The delicate balance between the methane homopause and cloud altitude in our models highlights the need for a more consistent treatment of cloud formation and disequilibrium chemistry.
	
	\item Our retrievals fit $\Kzz$ mainly based on the $3.4 \um$ methane feature. In the case of high-altitude clouds, a higher $\Kzz$ is preferred as it can mix additional material above the cloud deck. Therefore, we find that a high value for $\Kzz$ results in a larger amplitude feature. This is counter-intuitive as cloud-free chemistry models show that strong mixing quenches the atmosphere more deeply, resulting in overall less \ce{CH4} in the quenched layers. 
	
	\item We find no significant contribution of \ce{NH3} at $3$ or $10.5 \um$ in our cloudy model spectra, although a detection has been reported in WASP-107b. While high-altitude clouds could shield the \ce{NH3}-rich layers, we do not find a substantial contribution of \ce{NH3} in cloud-free spectra either.
\end{itemize}

\noindent Our study shows that hybrid retrievals can be a useful tool for analysing high-quality data.
However, the use of forward models introduces additional complexity and necessitates a thorough analysis before they can be incorporated into a fitting routine.

\begin{acknowledgements}
	We thank Paul Mollière for assisting with the use of petitRADTRANS, and Shang-Min Tsai for insightful discussions related VULCAN and our results.
	We also thank Mats Esseldeurs for his valuable advice throughout the development of this research.
	Finally, we thank the anonymous referee for their insightful comments that highly improved this work.
	T.K. and L.D. acknowledge funding from the KU Leuven Interdisciplinary Grant ESCHER (IDN/19/028).
	L.H. and L.D. acknowledge funding from the European Union H2020-MSCA-ITN-2019 under Grant no. 860470 (CHAMELEON).
	R.B. acknowledges support from the FNWI Origins investment incentive.
	K.H. acknowledges funding from the BELSPO FED-tWIN research program STELLA (Prf-2021-022) and the FWO research grant G014425N.
	V.C. thanks the Belgian F.R.S.-FNRS, and the Belgian Federal Science Policy Office (BELSPO) for the provision of financial support in the framework of the PRODEX Programme of the European Space Agency (ESA) under contract number 4000142531.

\end{acknowledgements}

\bibliographystyle{aa}
\bibliography{wasp107b}

\begin{appendix}

\onecolumn

\section{Additional VMR profiles} \label{appendix: chemistry of other molecules}

\begin{figure*}[h!]
	\centering
	\includegraphics[width=17cm]{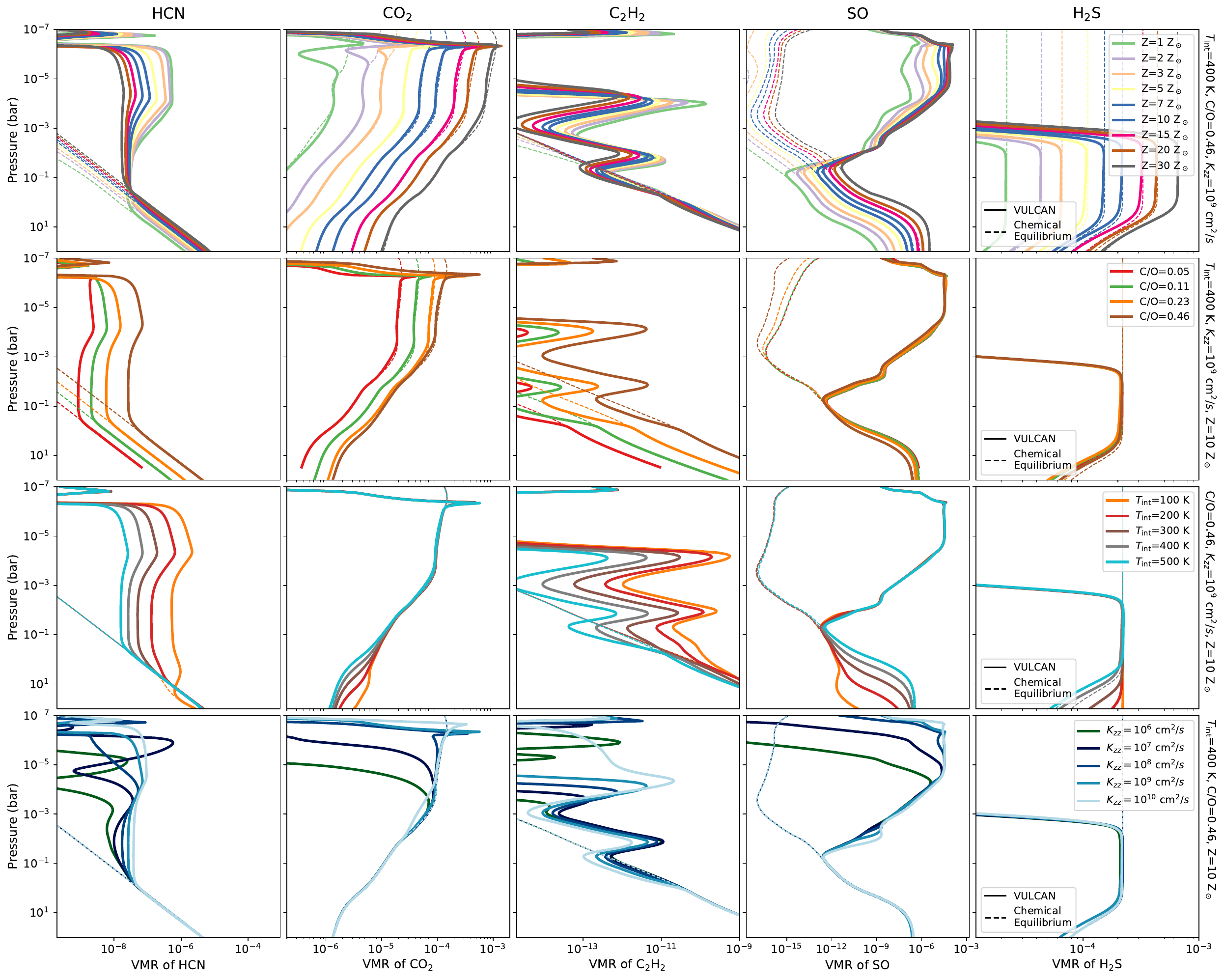}
	\caption{Same as Fig. \ref{fig: Chem_grid_overview_CH4_SO2_NH3} for \ce{HCN}, \ce{CO2}, \ce{C2H2}, \ce{SO}, and \ce{H2S}.}
	\label{appendixfig: Chemistry_grid_plot_HCN_CO2_C2H2_SO_H2S}
\end{figure*}

\FloatBarrier
\twocolumn

\begin{figure*}
	\centering
	\includegraphics[width=17cm]{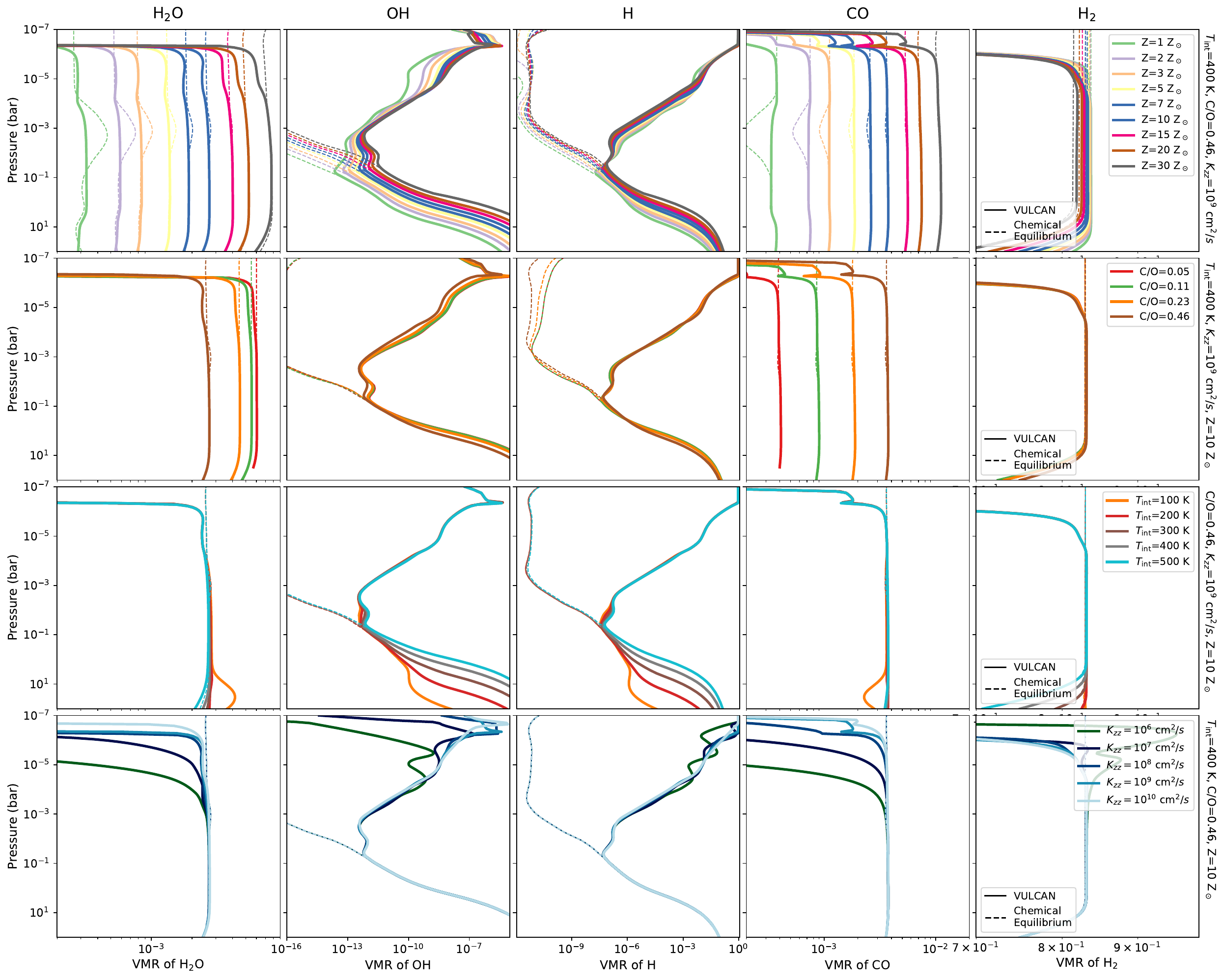}
	\caption{Same as Fig. \ref{fig: Chem_grid_overview_CH4_SO2_NH3} for \ce{H2O}, \ce{OH}, \ce{H}, \ce{CO}, and \ce{H2}.}
	\label{appendixfig: Chemistry_grid_plot_H2O_OH_H_H2_CO}
\end{figure*}

\begin{figure*}
	\centering
	\includegraphics[width=12cm]{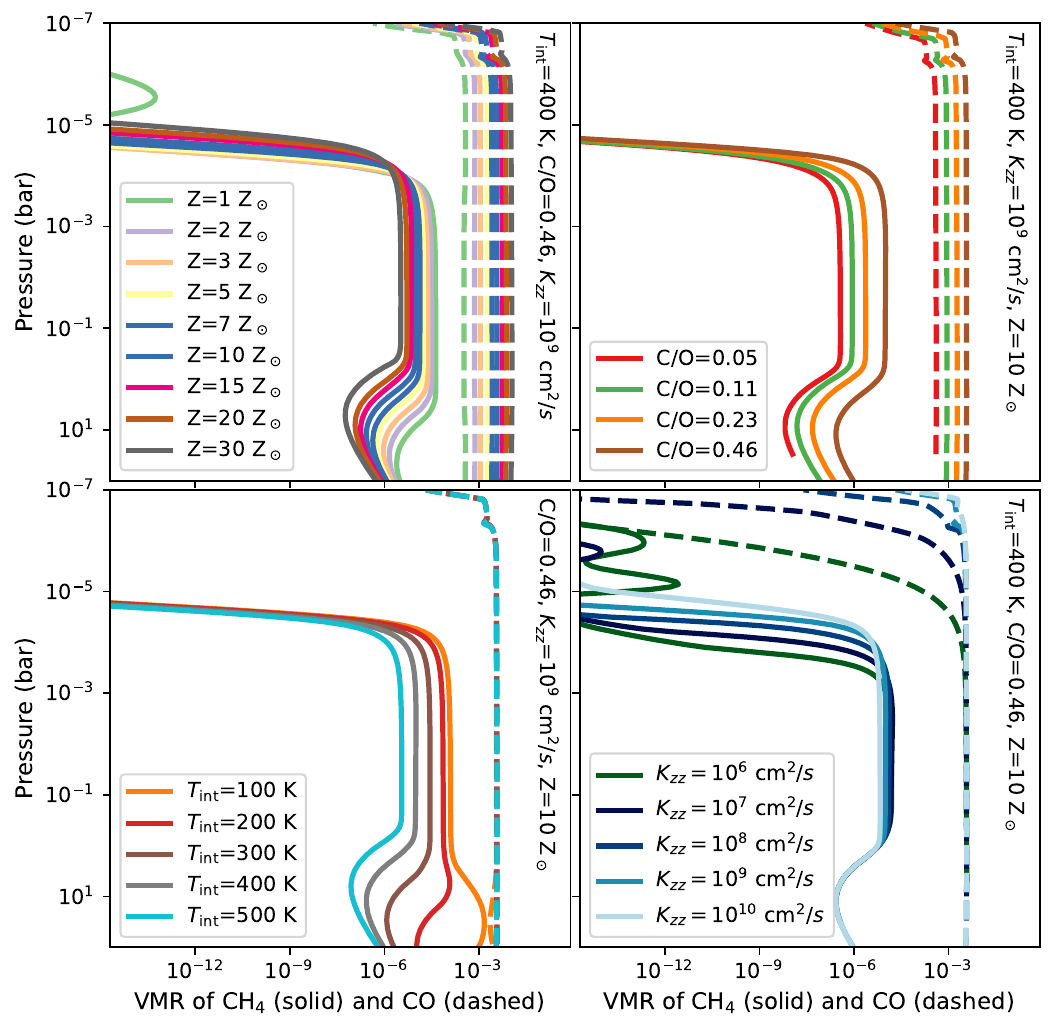}
	\caption{VMR profiles of \ce{CO} and \ce{CH4} in our grid of disequilibrium chemistry models of WASP-107b.}
	\label{appendixfig: 2x2_chemistryplotCH4_CO}
\end{figure*}

\begin{figure*}
	\centering
	\includegraphics[width=12cm]{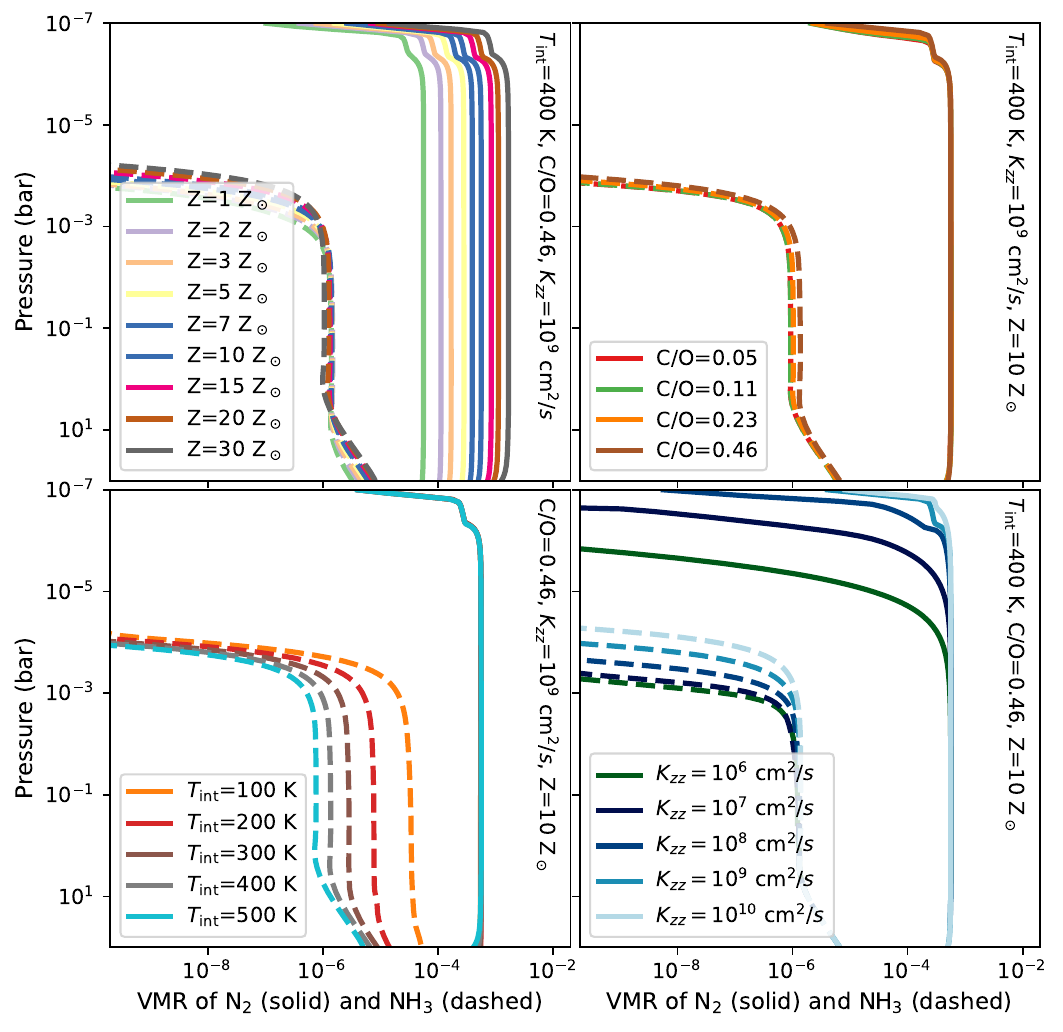}
	\caption{VMR profiles of \ce{N2} and \ce{NH3} in our grid of disequilibrium chemistry models of WASP-107b. }
	\label{appendixfig: 2x2_chemistryplotN2_NH3}
\end{figure*}

\FloatBarrier

\onecolumn

\section{Corner plots of Cases 1 and 10} 

\begin{figure*}[h!]
	\centering
	\includegraphics[width=17cm]{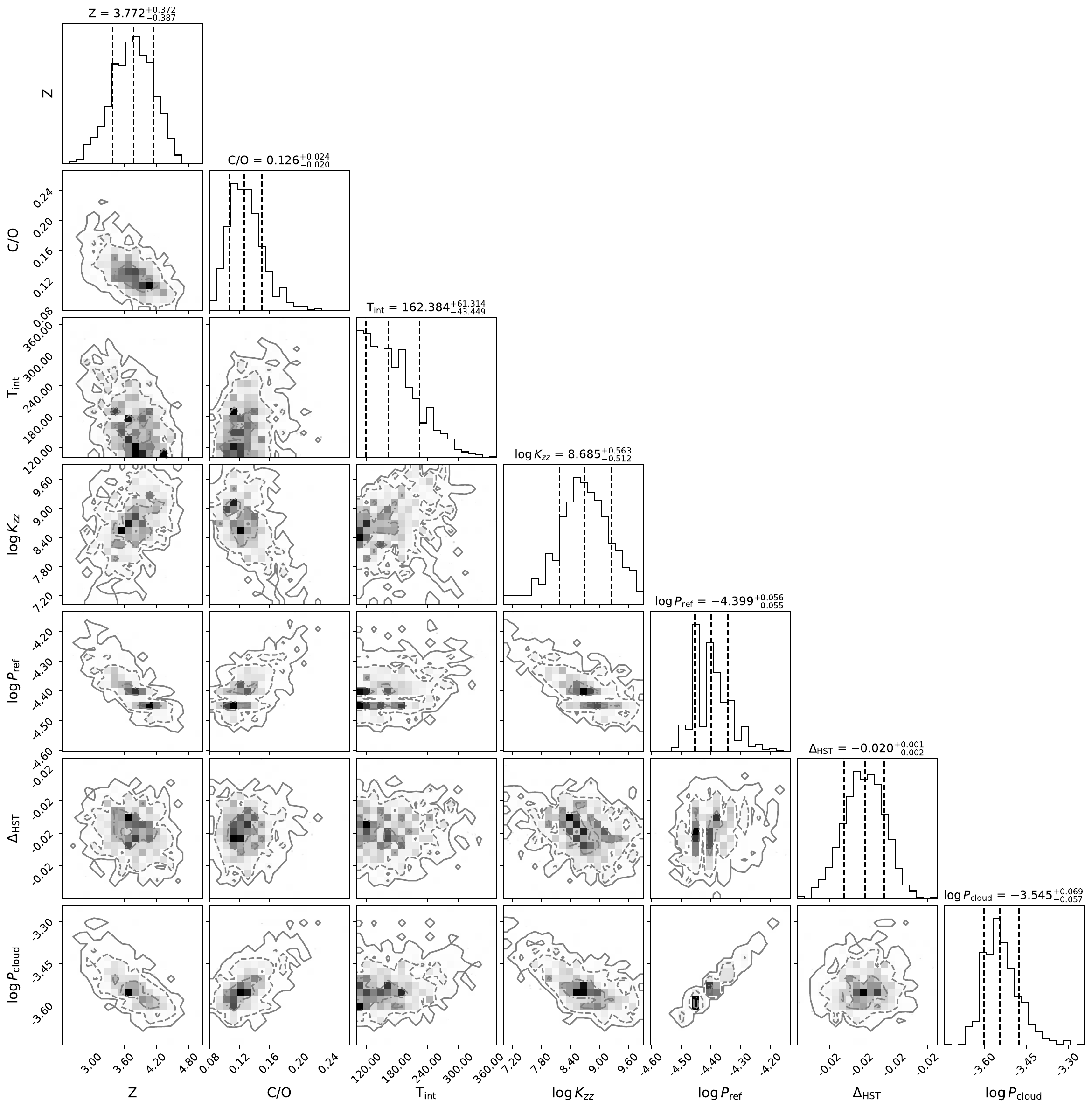}
	\caption{Corner plot for Case 1.}
	\label{fig: Ultranest-HSTNIRCam_FR_GC_cornetplot}
\end{figure*}

\FloatBarrier
\twocolumn

\begin{figure}
	\centering
	\includegraphics[width=17cm]{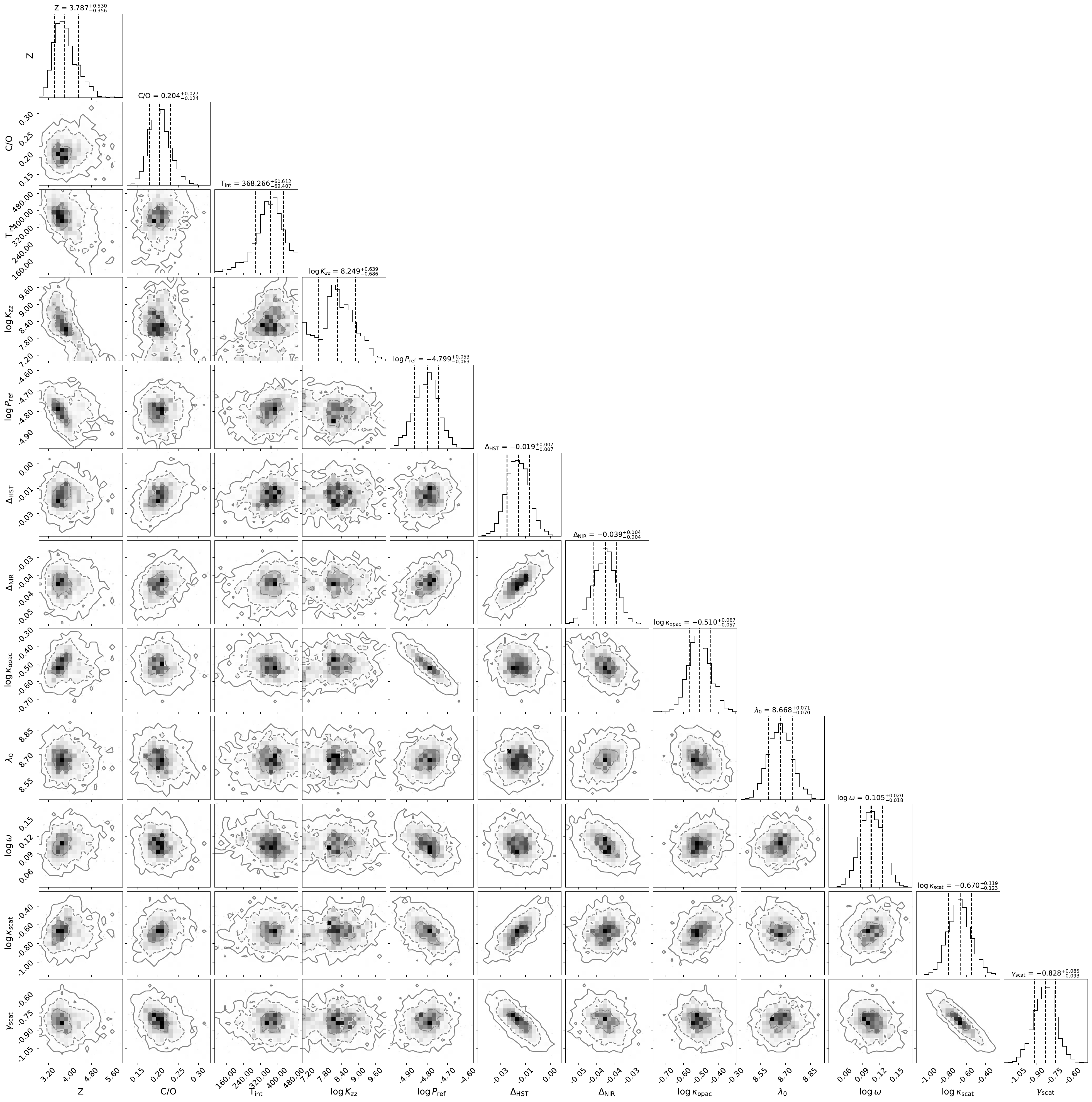}
	\caption{Corner plot for Case 10.}
	\label{fig: CornerPlot_LW}
\end{figure}

\FloatBarrier

\onecolumn
\section{Best-fit spectra of Cases 15, 16, and 17}

\begin{figure*}[h!]
	\centering
	\includegraphics[width=17cm]{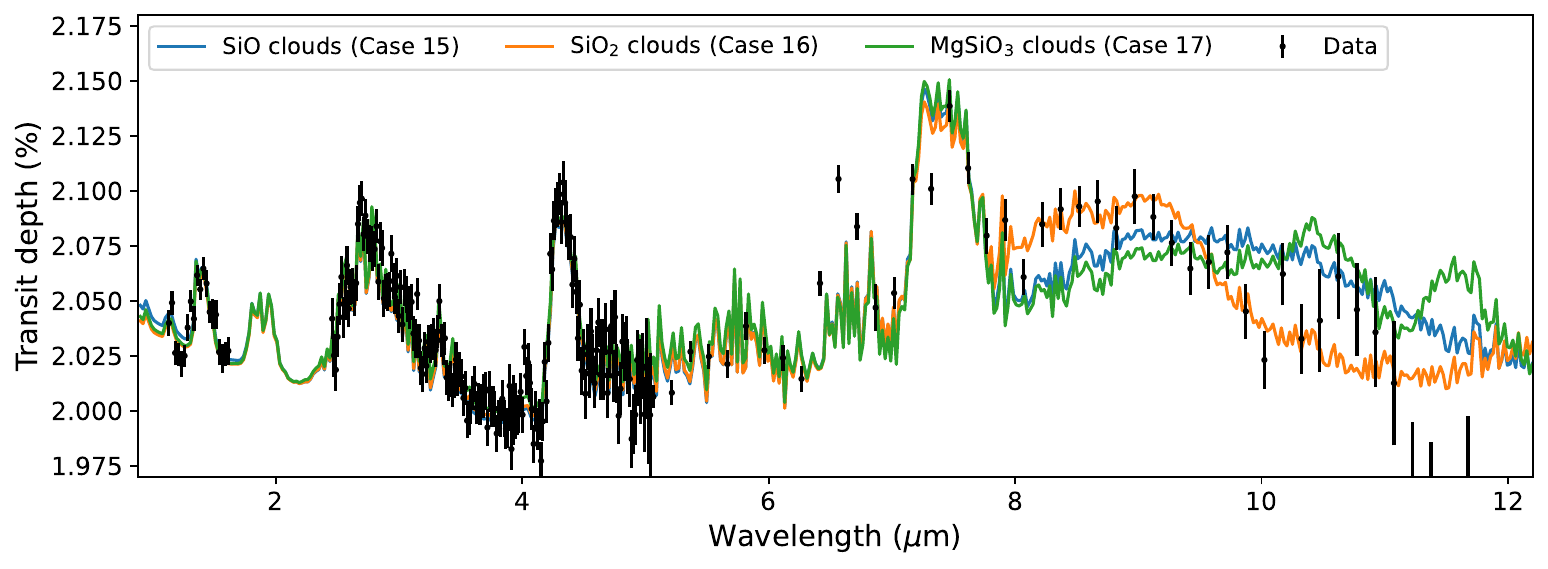}
	\caption{Best-fit spectra for Cases 15, 16, and 17 in which silicate condensates were fitted to the observations.}
	\label{appendixfig: Silicate_clouds}
\end{figure*}



\section{Tabulated retrieval outcomes}
\twocolumn


\begin{sidewaystable*}
	\begin{center}
		\centering
		\caption{\label{tab: OverviewUltranest} Summary of the retrievals with grey clouds with the nominal model (Case 1) listed first.
		}
		\begin{tabular}{lc@{\hskip 0.3in}cccccccc}
			\hline\hline\rule[0mm]{0mm}{5mm}
			& Case 1 & Case 2  & Case 3 & Case 4 & Case 5 & Case 6 & Case 7 & Case 8 & Case 9 \\
			& (Fig. \ref{fig: Ultranest-HSTNIRCam_FR_GC_bestfit}) &&&&&&&& \\

			&&  \multicolumn{5}{c}{\dotfill \textit{Sensitivity analysis}\dotfill} &  & \multicolumn{2}{c}{\dotfill \textit{Data information}\dotfill} \\
			&&&&&&&&& \\
			
			Tag & \textit{`nominal`} & \textit{`fix Z`} & \textit{`fix C/O`} & \textit{`fix $\Tint$`} & \textit{`fix $\Kzz$`} & \textit{`fix $\Pbase$`} & \textit{`free $\gamma$`} & \textit{`HST`} & \textit{`NIRcam`} \\
			Clouds & Grey & Grey & Grey & Grey & Grey & Grey  & Grey & Grey & Grey \\	
			Scattering &  No & No & No & No & No & No & Yes & No & No \\
			Datasets $^{(a)}$ & HN & HN & HN & HN & HM & HN & HN & H & N \\
			
			\hline\hline\rule[0mm]{0mm}{5mm}Parameter &&&&&&&&& \\
			\hline\rule[0mm]{0mm}{5mm}Z  ($\Zsun$)	$\in \left[1, 30 \right]$	& $3.77_{-0.39}^{+0.37}$ & 10 (fix) 				& $2.71_{-0.17}^{+0.22}$ & $2.76_{-0.49}^{+1.04}$ & $3.92_{-0.41}^{+0.36}$ & $1.74_{-0.08}^{+0.08}$ & $4.85_{-0.46}^{+0.49}$ & $17.57_{-10.52}^{+7.90}$ & $4.08_{-0.39}^{+0.32}$ \\
			&&&&&&&&& \\
			
			C/O $\in \left[0.05, 0.46 \right]$		& $0.13_{-0.02}^{+0.02}$ & $0.11_{-0.01}^{+0.01}$ 	& 0.46 (fix) 			& $0.20_{-0.06}^{+0.05}$ & $0.13_{-0.02}^{+0.02}$ & $0.22_{-0.02}^{+0.03}$ & $0.15_{-0.03}^{+0.03}$ & $0.26_{-0.13}^{+0.13}$ 	& $0.09_{-0.01}^{+0.02}$   \\
			&&&&&&&&& \\
			
			$\Tint$ (K) $\in \left[100, 500\right]$ & $163_{-44}^{+62}$		& $264_{-110}^{+172}$ 		& $490_{-15}^{+7}$ 		& 500 (fix) 			& $235_{-29}^{+38}$ & $489_{-15}^{+8}$ & $229_{-82}^{+85}$ & $313_{-148}^{+126}$		 & $156_{-40}^{+56}$  \\
			&&&&&&&&& \\	
			
			$\log \Kzz$ (cgs) $\in \left[7, 10\right]$  & $8.68_{-0.51}^{+0.56}$ & $7.05_{-0.04}^{+0.08}$ & $7.18_{-0.13}^{+0.22}$ & $7.31_{-0.23}^{+0.57}$ & 10 (fix) 			& $7.18_{-0.13}^{+0.21}$ & $8.48_{-0.47}^{+0.60}$ & $8.53_{-1.01}^{+0.99}$ & $8.64_{-0.53}^{+0.49}$  \\
			&&&&&&&&& \\

			\hline
			&&&&&&&&& \\
			
			$\log \Pref$ (bar) $\in  \left[-5, -2\right]$ 	& $-4.40_{-0.05}^{+0.06}$ & $-4.72_{-0.01}^{+0.01}$ & $-4.08_{-0.05}^{+0.02}$ & $-4.17_{-0.16}^{+0.10}$ & $-4.46_{-0.05}^{+0.06}$ & $-3.89_{-0.01}^{+0.01}$  & $-4.64_{-0.06}^{+0.05}$ & $-4.95_{-0.20}^{+0.29}$ & $-4.35_{-0.03}^{+0.05}$  \\
			&&&&&&&&& \\
			
			$\offsetHST$ (\%) $\in \left[-0.1, 0.1\right]$ & $-0.02_{-0.00}^{+0.00}$ & $-0.02_{-0.00}^{+0.00}$ & $-0.02_{-0.00}^{+0.00}$ & $-0.02_{-0.00}^{+0.00}$ & $-0.02_{-0.00}^{+0.00}$ & $-0.02_{-0.00}^{+0.00}$ & $0.02_{-0.00}^{+0.00}$ && \\
			&&&&&&&&& \\
			
			
			\hline
			&&&&&&&&& \\
			
			$\log \Pbase$ (bar) $\in \left[-5, 0\right]$ & $-3.54_{-0.06}^{+0.07}$ & $-3.86_{-0.03}^{+0.02}$ & $-3.23_{-0.05}^{+0.04}$ & $-3.33_{-0.16}^{+0.11}$ & $-3.60_{-0.06}^{+0.06}$ & -3 (fix) & $-2.03_{-0.66}^{+0.69}$ & $-4.21_{-0.20}^{+0.25}$ & $-3.46_{-0.05}^{+0.06}$  \\
			&&&&&&&&& \\
			
			$\log \kappascatt$ (cgs) $\in  \left[-2, 2\right]$ 		&  &  &   & &   & & $-0.61_{-0.11}^{+0.12}$ &&  \\
			&&&&&&&&& \\
			
			$\gammascatt \in \left[-4, 0\right]$      		&  &  &	 & 	& & & $-0.75_{-0.09}^{+0.09}$  && \\
			&&&&&&&&& \\
			
			\hline 
			&&&&&&&&& \\
			
			$\chi^2_{\rm red}$ 	& $2.18$ 	& $2.75$ & $2.49$	& $2.26$	& $2.20$ & $2.43$ & $1.75$	 & $2.07$ & $2.06$  	\\
			&&&&&&&&& \\
			
			BIC		&	 $\num{-1050}$ & $\num{-946}$	& $\num{-995}$	& $\num{-1038}$& $\num{-1050}$ & $\num{-1005}$ & $\num{-1123}$ & $\num{-125}$ &  $\num{-948}$\\
			&&&&&&&&& \\

			\hline\rule[0mm]{0mm}{5mm}
			
		\end{tabular}
	\end{center}
	{$^{(a)} $ Models were fitted to HST WFC3 (H) and JWST NIRCam (N) data. }
\end{sidewaystable*}



\begin{table*}
	\begin{center}
		\centering
		\caption{\label{tab: OverviewUltranest_Gaussianclouds}Summary of the retrievals with non-grey `Gaussian` clouds with the nominal model (Case 10) listed first.
		}
		\begin{tabular}{lc@{\hskip 0.3in}cccc}
			\hline\hline\rule[0mm]{0mm}{5mm}
			&  Case 10 & Case 11  & Case 12 & Case 13 & Case 14 \\
			& (Fig. \ref{fig: Bestfit_LW}) &&&& \\
			
			
			Tag & \textit{`nominal`} & \textit{`Fix Z`} & \textit{`Fix C/O`} & \textit{`No $\gamma$`} &  \textit{`MIRI`}  \\	
			Clouds & Gaussian & Gaussian & Gaussian & Gaussian & Gaussian \\	
			Scattering &  Yes & Yes & Yes & No & No  \\
			Datasets $^{(a)}$ & HNM & HNM & HNM & HNM & M \\
			
			\hline\hline\rule[0mm]{0mm}{5mm}
			Parameter  & &&&& \\
			\hline\rule[0mm]{0mm}{5mm}
			
			Z  ($\Zsun$)	$\in \left[1, 30 \right]$			& $3.78_{-0.35}^{+0.53}$ & 10 (fix) & $5.10_{-0.28}^{+0.40}$ & $3.43_{-0.29}^{+0.30}$ &  $3.16_{-0.43}^{+0.78}$  \\
			&&&&& \\
			
			C/O $\in \left[0.05, 0.46 \right]$		& $0.20_{-0.02}^{+0.03}$ & $0.42_{-0.03}^{+0.03}$ & 0.46 (fix) & $0.15_{-0.02}^{+0.02}$ & $0.12_{-0.06}^{+0.10}$  \\
			&&&&& \\
			
			$\Tint$ (K) $\in \left[100, 500\right]$		& $369_{-70}^{+60}$	& $185_{-56}^{+81}$ & $459_{-43}^{+28}$ & $206_{-71}^{+76}$ &  $395_{-163}^{+79}$  \\
			&&&&& \\	
			
			$\log \Kzz$ (cgs) $\in \left[7, 10\right]$ 	& $8.25_{-0.68}^{+0.64}$ & $7.02_{-0.02}^{+0.04}$ &  $7.32_{-0.21}^{+0.28}$ & $7.72_{-0.46}^{+0.52}$ & $7.49_{-0.37}^{+1.01}$ \\
			&&&&& \\

			\hline
			&&&&& \\
			
			$\log \Pref$ (bar) $\in  \left[-5, -2\right]$ 		& $-4.80_{-0.06}^{+0.05}$ & $-5.00_{-0.00}^{+0.01}$ & $-4.75_{-0.06}^{+0.05}$ & $-4.65_{-0.05}^{+0.05}$ &  $-4.49_{-0.23}^{+0.12}$  \\
			&&&&& \\
			
			$\offsetHST$ (\%) $\in \left[-0.1, 0.1\right]$		& $-0.02_{-0.01}^{+0.01}$ & $0.01_{-0.01}^{+0.01}$ & $0.00_{-0.01}^{+0.01}$ & $-0.07_{-0.00}^{+0.00}$ &  \\
			&&&&& \\
			
			$\offsetNIR$ (\%) $\in \left[-0.1, 0.1\right]$ 		& $-0.04_{-0.00}^{+0.00}$ & -$0.03_{-0.00}^{+0.00}$ & $-0.03_{-0.00}^{+0.00}$ & $-0.05_{-0.00}^{+0.00}$ &  \\
			&&&&& \\
			
			\hline
			&&&&& \\
			
			$\log \Pbase$ (bar) $\in \left[-5, 0\right]$    	&  & & & $-3.38_{-0.06}^{+0.05}$ &  $-3.09_{-0.19}^{+0.16}$  \\
			&&&&& \\
			
			$\log \kappascatt$ (cgs) $\in  \left[-2, 2\right]$ 		& $-0.67_{-0.12}^{+0.12}$ & $-0.03_{-0.10}^{+0.10}$ & $-0.42_{-0.11}^{+0.11}$ &		  &    \\
			&&&&& \\
			
			$\gammascatt \in \left[-4, 0\right]$        		& $-0.83_{-0.09}^{+0.09}$ & $-1.23_{-0.09}^{+0.09}$ &	$-1.08_{-0.08}^{+0.08}$ &		&  \\
			&&&&& \\

			\hline
			&&&&& \\
			
			$\log \kappaWelbanks$ (cgs) $\in  \left[-1, 1\right]$ 	& $-0.51_{-0.06}^{+0.07}$ & $-0.41_{-0.03}^{+0.03}$ & $-0.54_{-0.06}^{+0.06}$ & $-0.61_{-0.05}^{+0.06}$ &  $-0.79_{-0.13}^{+0.24}$ \\
			&&&&& \\
			
			$\lambda_0$ ($\um$)  $\in \left[5, 12\right]$ 			& $8.67_{-0.07}^{+0.07}$ & $8.62_{-0.07}^{+0.07}$ & $8.59_{-0.06}^{+0.07}$ & $8.68_{-0.07}^{+0.08}$ &  $8.72_{-0.14}^{+0.11}$  \\
			&&&&& \\
			
			$\log \omega \in  \left[-1, 0.5\right]$ 		& $0.11_{-0.02}^{+0.02}$ & $0.13_{-0.02}^{+0.02}$ &
			$0.10_{-0.02}^{+0.02}$ & $0.09_{-0.02}^{+0.02}$ &  $0.13_{-0.03}^{+0.04}$  \\
			&&&&& \\

			\hline 
			&&&&& \\
			
			$\chi^2_{\rm red}$ & $2.19$ & $ 2.56$ & $ 2.34$ & $2.60$ & $4.53$  \\
			&&&&& \\
			
			BIC & $\num{-1263}$ & $\num{-1181}$ & $\num{-1232}$ & $\num{-1173}$ & $\num{-132}$\\
			&&&&& \\
			
			\hline\rule[0mm]{0mm}{5mm}
		\end{tabular}
	\end{center}
	{$^{(a)} $ Models were fitted to HST WFC3 (H), JWST NIRCam (N), and JWST MIRI/LRS (M) data. }
\end{table*}


\begin{table*}
	\begin{center}
		\centering
		\caption{\label{tab: OverviewUltranest_Silicateclouds}Summary of the retrievals with non-grey silicate clouds.
		}
		\begin{tabular}{lccc}
			\hline\hline\rule[0mm]{0mm}{5mm}
			&  Case 15 & Case 16 & Case 17 \\
			& & & \\
			
			
			Tag & \textit{`\ce{SiO}`} & \textit{`\ce{SiO2}`} & \textit{`\ce{MgSiO3}`} \\	
			Clouds &  \ce{SiO} & \ce{SiO2} & \ce{MgSiO3} \\	
			Scattering &  - & - & - \\
			Datasets $^{(a)} $ & HNM & HNM & HNM \\
			
			\hline\hline\rule[0mm]{0mm}{5mm}
			Parameter  &&& \\
			\hline\rule[0mm]{0mm}{5mm}
			
			Z  ($\Zsun$)	$\in \left[1, 30 \right]$			 & $3.81_{-0.20}^{+0.21}$ & $3.59_{-0.24}^{+0.22}$ & $3.89_{-0.17}^{+0.18}$ \\
			&&& \\
			
			C/O $\in \left[0.05, 0.46 \right]$		&  $0.15_{-0.02}^{+0.02}$ & $0.16_{-0.02}^{+0.02}$ & $0.14_{-0.02}^{+0.02}$  \\
			&&& \\
			
			$\Tint$ (K) $\in \left[100, 500\right]$		& $170_{-49}^{+66}$ & $201_{-66}^{+79}$ & $144_{-32}^{+57}$ \\
			&&& \\	
			
			$\log \Kzz$ (cgs) $\in \left[7, 10\right]$ 	& $8.43_{-0.37}^{+0.49}$ & $8.32_{-0.36}^{+0.53}$ & $8.59_{-0.32}^{+0.36}$ \\
			&&& \\

			\hline
			&&& \\
			
			$\log \Pref$ (bar) $\in  \left[-5, -2\right]$ 	 & $-4.64_{-0.04}^{+0.04}$ & $-4.60_{-0.04}^{+0.04}$ & $-4.61_{-0.03}^{+0.04}$ \\
			&&& \\
			
			$\offsetHST$ (\%) $\in \left[-0.1, 0.1\right]$	&  $-0.02_{-0.00}^{+0.00}$ & $-0.03_{-0.00}^{+0.00}$ & $-0.02_{-0.00}^{+0.00}$\\
			&&& \\
			
			$\offsetNIR$ (\%) $\in \left[-0.1, 0.1\right]$ 		&  $-0.03_{-0.00}^{+0.00}$ & $-0.03_{-0.00}^{+0.00}$ & $-0.03_{-0.00}^{+0.00}$ \\
			&&& \\
			
			\hline
			&&& \\
			
			$\log \Pbase$ (bar) $\in \left[-5, 0\right]$    	&   $-2.71_{-0.42}^{+0.45}$ & $-3.10_{-0.16}^{+0.17}$ & $-2.89_{-0.33}^{+0.37}$ \\
			&&& \\
			
			$\log \kappascatt$ (cgs) $\in  \left[-2, 2\right]$ 		  & & & \\
			&&& \\
			
			$\gammascatt \in \left[-4, 0\right]$        		& & & \\
			&&& \\

			\hline
			&&& \\

			$\log \Xbase \in  \left[-6, -2\right]$ 			& $-2.09_{-0.74}^{+0.78}$ & $-1.68_{-0.45}^{+0.49}$ & $-2.00_{-0.65}^{+0.67}$\\
			&&& \\
			
			$\fsed \in \left[0, 10\right]$ 				& $1.79_{-0.16}^{+0.17}$ & $2.91_{-0.18}^{+0.20}$ &$2.03_{-0.19}^{+0.23}$  \\
			&&& \\
			
			$\log \Kzzcloud$ $ \in \left[7, 12\right]$ 		& $7.56_{-0.09}^{+0.09}$ & $7.30_{-0.08}^{+0.08}$ & $7.84_{-0.10}^{+0.10}$ \\
			(cgs) &&& \\
			
			\hline 
			&&& \\
			
			$\chi^2_{\rm red}$ &  $2.88$ & $2.60$ & $3.28$ \\
			&&& \\
			
			BIC  & $\num{-1107}$ & $\num{-1173}$ & $\num{-1014}$ \\
			&&& \\
			
			\hline\rule[0mm]{0mm}{5mm}
		\end{tabular}
	\end{center}
	{$^{(a)} $ Models were fitted to HST WFC3 (H), JWST NIRCam (N), and JWST MIRI/LRS (M) data. }
\end{table*}


\begin{table*}
	\begin{center}
		\centering
		\caption{\label{tab: grid6Comparison}
			Summary of the retrievals with a slightly cooler grid {(r$_{\rm st}  = 0.5$)} of forward models compared to the previous models presented in this study.
		}
		\begin{tabular}{lcc}
			\hline\hline\rule[0mm]{0mm}{5mm}
			& Case 18 & Case 19 \\
			& & (Fig. \ref{fig: Grid6Comparison}) \\
			&& \\
			Tag & \textit{`r$_{\rm st}=0.5$`} & \textit{`r$_{\rm st}=0.5$`}  \\
			& (set-up Case 1) & (set-up Case 10) \\
			
			Clouds & Grey & Gaussian \\	
			Scattering & No & Yes \\
			Datasets $^{(a)} $ & HN & HNM \\
			
			\hline\hline\rule[0mm]{0mm}{5mm}
			Parameter $^{(b)} $ & & \\
			\hline\rule[0mm]{0mm}{5mm}
			
			Z ($\Zsun$) & $3.92_{-0.36}^{+0.34}$ & $4.49_{-0.27}^{+0.32}$ \\
			&& \\
			
			C/O & $0.07_{-0.01}^{+0.01}$ & $0.11_{-0.02}^{+0.03}$  \\
			&& \\
			
			$\Tint$ (K) & $454_{-36}^{+29}$ & $482_{-21}^{+13}$ \\
			&& \\	
			
			$\log \Kzz$ (cgs) &  $7.75_{-0.54}^{+0.62}$ & $8.03_{-0.77}^{+0.78}$ \\
			&&\\

			\hline
			&& \\
			
			$\log \Pref$ (bar) & $-4.13_{-0.04}^{+0.04}$ & $-4.69_{-0.05}^{+0.06}$ \\
			&& \\Tab
			
			$\offsetHST$ (\%) & $-0.02_{-0.00}^{+0.00}$ & $-0.02_{-0.01}^{+0.01}$\\
			&& \\
			
			$\offsetNIR$ (\%)  &  & $-0.04_{-0.00}^{+0.00}$ \\
			&& \\
			
			\hline
			&& \\
			
			$\log \Pbase$ (bar) & $-3.12_{-0.04}^{+0.05}$ &  \\
			&& \\
			
			$\log \kappascatt$ (cgs) & & $-0.84_{-0.14}^{+0.13}$ \\
			&& \\
			
			$\gammascatt$ & & $-1.00_{-0.11}^{+0.10}$ \\
			&& \\

			\hline
			&& \\
			
			$\log \kappaWelbanks$ (cgs) &  & $-0.74_{-0.07}^{+0.06}$ \\
			&& \\
			
			$\lambda_0$ ($\um$) &  & $8.60_{-0.07}^{+0.07}$ \\
			&& \\
			
			$\log \omega$ &  & $0.09_{-0.02}^{+0.02}$  \\
			&& \\
			
			\hline
			&& \\
			
			$\chi^2_{\rm red}$ & $2.23$ & $2.31$ \\
			&&\\
			
			BIC & $\num{-1040}$ & $\num{-1236}$ \\
			&  & \\
			
			\hline 
			
		\end{tabular}
	\end{center}
	{$^{(a)} $ Models were fitted to HST WFC3 (H), JWST NIRCam (N), and JWST MIRI/LRS (M) data. } \\ 
	{$^{(b)}$ The prior ranges are the same as in Tables \ref{tab: OverviewUltranest} and \ref{tab: OverviewUltranest_Gaussianclouds}.}
	
\end{table*}

\end{appendix}

\end{document}